\input psfig.sty 
%
%
%

\ifx\mnmacrosloaded\undefined \input mn\fi

%

\newif\ifAMStwofonts
\AMStwofontstrue 

\ifCUPmtplainloaded \else
  \NewTextAlphabet{textbfit} {cmbxti10} {}
  \NewTextAlphabet{textbfss} {cmssbx10} {}
  \NewMathAlphabet{mathbfit} {cmbxti10} {} 
  \NewMathAlphabet{mathbfss} {cmssbx10} {} 
  \ifAMStwofonts
    \NewSymbolFont{upmath} {eurm10}
    \NewSymbolFont{AMSa} {msam10}
    \NewMathSymbol{\upi}     {0}{upmath}{19}
    \NewMathSymbol{\umu}     {0}{upmath}{16}
    \NewMathSymbol{\upartial}{0}{upmath}{40}
    \NewMathSymbol{\leqslant}{3}{AMSa}{36}
    \NewMathSymbol{\geqslant}{3}{AMSa}{3E}

  \else     \def\umu{\mu}
    \def\upi{\pi}
    \def\upartial{\partial}
  \fi
\fi


\pageoffset{-2.5pc}{0pc}

\loadboldmathnames

 \Referee   


\onecolumn        
\pagerange{}    
\pubyear{2005}
\volume{}

\begintopmatter  

\title{Multicomponent Decompositions for a Sample of S0 galaxies}
\author{Eija Laurikainen and Heikki Salo}
\affiliation{Division of Astronomy, Dept. of Physical. Sciences, University of Oulu, FIN-90014, Finland}

and

\author{Ronald Buta}
\affiliation{Department of Physics and Astronomy, Box 870324, Univ.of Alabama, Tuscaloosa, AL 35487, USA}

\shortauthor{E. Laurikainen, H. Salo, R. Buta}
\shorttitle{decompositions for S0s}



email:eija.laurikainen@oulu.fi

\abstract {We have estimated the bulge-to-total ($B/T$) light ratios in 
the $K_s$-band for a sample of 24 S0, S0/a and Sa galaxies by applying
a 2-dimensional multicomponent decomposition method. 
For the disk an exponential function is used, the bulges are fitted by a 
S\'ersic's
$R^{1/n}$ function and the bars and ovals are described either by a S\'ersic 
or a Ferrers function.  
In order to avoid non-physical solutions, preliminary characterization 
of the structural components is made by inspecting the radial profiles 
of the orientation parameters and the  
low azimuthal wavenumber Fourier amplitudes and phases. 
In order to identify also the inner structures,
unsharp masks were created: previously undetected inner spiral
arms were found in NGC 1415 and marginally in NGC 3941. Most importantly,     
we found that S0s have a mean $<B/T>_K$-ratio of 0.24 $\pm$ 0.11, 
which is significantly smaller than the mean $<B/T>_R$ = 0.6 generally 
reported in the literature. Also, the surface brightness profiles of the 
bulges in S0s were found to be more exponential-like than generally assumed, 
the mean shape parameter of the bulge being $<n>$ = 2.1 $\pm$ 0.7.
We did not find examples of barred S0s lacking the disk component, 
but we found some galaxies (NGC 718, NGC 1452, NGC 4608) having a 
non-exponential disk in 
the bar region. To our knowledge our study is the first attempt to
apply a multicomponent decomposition method for a moderately sized sample 
of early-type disk galaxies.
}


\maketitle  

\section{Introduction}

In the present view galaxy evolution has two domains, rapid processes
related to hierarchical clustering and merging leading to formation of
the main structural components of galaxies (Eggen, Lynden-Bell $\&$ Sandage;
Toomre 1977; Firmani $\&$ Avila-Rees 2003), and slow secular evolution 
occurring in later phases of galaxy evolution (Kormendy $\&$ Kennicutt 2004,
hereafter KK04). The slow internal evolutionary processes are assumed to 
be important for
spiral galaxies, but less likely in S0s, where externally induced ram 
pressure stripping might play a more important role, possibly transforming 
spiral galaxies into S0s by a loss of gas content (Bekki, Warrick $\&$ 
Yasuhiro 2002). 

One of the main reasons to believe that internal secular evolution 
might be important for spirals, but not for S0s, is the fact that spirals 
have small exponential 
bulges or no bulges at all (Carollo et al. 1997; Carollo 1999). As discussed
by KK04, the lack of bulges in some spirals probably means that they 
have not suffered any major mergers during their lifetime. This indicates that
in principle there has been enough time for relatively isolated spirals 
to develop bulges via secular evolutionary processes. On the other hand, 
S0s are found to have massive bulges (Simien $\&$ de Vaucouleurs 1986) 
with surface brightness profiles similar to those in elliptical galaxies,
(Andredakis, Peletier $\&$ Balcells 1995, hereafter APB95; Gadotti $\&$ de Souza 2003, 
hereafter GS03), 
which are easily produced in mergers of two massive disk galaxies
(Bekki 1995). However, the properties of S0s are still poorly known. 
For example, there is recent evidence showing that the bulges in
many S0s might be rotationally supported (Erwin, Beckman $\&$ Beltran 2004) 
in a similar manner 
as the pseudo-bulges in spirals. S0s are also found to be complex
systems sometimes having many bars and ovals in the same galaxy 
(Peng et al. 2002; Erwin et al. 2003), which challenges any
simple structural decompositions made for these galaxies.  

One of the puzzles for S0s is that they have weak disks, but at the same time 
also strong bars in many cases. If these galaxies are 
bulge dominated as generally assumed, it would be difficult to explain the 
bar formation by a global instability in a cool 
disk (Binney $\&$ Tremaine 1987; Sellwood 2000). However, this paradox 
is resolved if it is assumed that there
is angular momentum exchange between the bar and the spheroidal component:
it was shown by Athanassoula (2002, 2003) that in the presence of a 
massive halo or a large massive bulge, angular momentum is emitted
in the inner disk resonances and absorbed in the resonances 
of the outer disk and halo. In this process an initially
weak bar grows stronger and the bar may also consume material that originally
belonged to the disk (Athanassoula 2003). In this picture it would be 
possible to have strong bars even in galaxies dominated by massive bulges. 
As an extreme case one would see a galaxy with a strong bar, but
no sign of the underlying disk, as suggested by GS03.

The above scenario for the formation of bars is an interesting approach 
to secular evolution of S0s,
but there are still many important observational properties related
to this picture that need to be re-investigated: for example, 
{\it do most S0s have large 
massive bulges} as generally assumed and, {\it do there 
exist strongly barred S0s without any sign of the disk component?} 
In this study we measure the bulge-to-total ($B/T$) light ratio for 24 
early-type disk galaxies, and 
analyze in detail NGC 4608, suggested to be a candidate galaxy with 
a strong bar lacking the disk component (GS03). 
We use a 2-dimensional multicomponent decomposition method,
in which the disk is described by an exponential function, the bulge  
by a S\'ersic's function, and for bars and ovals either  
a Ferrers or a S\'ersic function is used. In order to find physically 
reasonable solutions the images are first inspected by studying the 
radial profiles 
of the orientation parameters, by calculating low azimuthal wavenumber
Fourier amplitudes and phases, and by creating unsharp masks.

\section{The sample and data reductions}

Our sample consists of 24 nearby S0-Sa galaxies ($-$3 $<$ T $<$ 1) 
having total magnitudes $B_T < $ 12.5 mag and inclinations less than $65^o$. 
This is part of our Near-IR S0 Survey (NIRS0S) of 170 S0-Sa galaxies 
selected to be comparable in size, total apparent magnitude, and number 
with the Ohio State University Bright Galaxy Survey (OSUBGS, Eskridge et 
al. 2002) for spirals, for which we have previously made similar
analysis as presented in this study (Buta, Laurikainen, Salo 2004; 
Laurikainen, Salo, Buta 2004; Laurikainen et al. 2004, hereafter LSBV04). 
The galaxies are mainly S0s, but some early-type spirals 
are included to ensure a small overlap with the OSU sample, and also
because galaxies tend to look earlier in the near-IR than in the optical.
The subsample of 24 galaxies 
was selected to have a large variety of bar and bulge sizes. 
The sample is listed in 
Table 1, where the morphological types are from the Third Reference 
Catalog of Bright Galaxies (de Vaucouleurs et al. 1991, RC3). 
Some of the galaxies have active nuclei, the activity types being taken from 
the NASA/IPAC Extragalactic Database (NED).  

We present high resolution $K_s$ and $B$-band observations carried out
at the 2.5 m Nordic Optical telescope (NOT) in La Palma in Jan 2003 
and Jan 2004.
The observations were made in good weather conditions. 
Full Width at Half Maximum (FWHM) values for the point spread function
were measured using several foreground stars and are listed in Table 1.
The average value is 1.1 arcsec.
The $K_s$-band observations were obtained using NOTCam, a 1024 x 1024 
detector array with a pixel size of 0.23 arcsec pixel$^{-1}$ and a field 
of view of 4 x 4 arcmin$^2$. The total on-source integration time was 
generally 1800 sec and the integration time in one position was
taken in snapshots of 20-30 sec. Owing to the high 
sky brightness in the near-IR, the sky fields were taken in two  
directions to avoid bright foreground stars and alternating between
the galaxy and the sky field in periods of one minute. In order to avoid 
interference patterns and hot pixels in the images, dithering of 5 arcsec 
was used in the galaxy field and a larger dithering was used in the sky field. 
Twilight flats only were used and were constructed 
by subtracting a low ADU-level image from a high ADU-level image.

All images were processed using the IRAF package routines \footnote{${}^{1}$} 
{IRAF is distributed by the National Optical Astronomy Observatories,
which are operated by the Association of Universities for Research in 
Astronomy, under cooperative agreement with the National Science Foundation}.
The reduction steps consisted of sky subtraction, combining the on-source
integrations, flat-fielding, cleaning the images, and transposing the images to
have North up and West on the right. Generally the best sky subtraction was
achieved using the temporally nearest sky frame for each science frame.
Pickup noise in some of the images was removed using a destriping method 
described by Buta $\&$ McCall (1999). Images were cleaned of foreground stars 
using DAOFIND to find the stars initially. Then a combination of point 
spread function (PSF) fitting and image editing (IMEDIT) was used to 
remove the stars.

Optical $B$-band observations were obtained using the 2048 x 2048 ALFOSC 
CCD, which has a field of view of 6.5 x 6.5 arcmin$^2$ and a pixel 
size of 0.19 arcsec pixel$^{-1}$. For the high surface brightness
centers of many S0s, the integrations were generally divided in a number 
of short exposures so that the total integration time was 1800 sec. As the 
CCD field was always larger than the galaxy size, no extra sky fields 
were obtained. Twilight flats were used, and the standard reduction steps 
(combining the images, bias subtraction, flat-fielding,
cleaning and transposing the images) were performed using the IRAF routines.
For the galaxies with both optical and near-IR observations,
the relative difference in the direction of the North in the sky
plane was checked between the $B$ and $K_s$-band images using the 
IRAF routine GEOMAP. 

\section{Identification of the structural components}

S0 galaxies generally have complicated morphological structures, which 
challenges the determination of their bulge-to-total ($B/T$) light ratios.
Besides bulges and disks, S0s may also have bars, inner disks and ovals,  
which are not always visible in the azimuthally averaged surface 
brightness profiles: if a component in question has a low surface 
brightness compared to the surface brightness of the bulge, it is easily 
overshadowed by the bulge. Before any reliable 
decompositions can be made, {\it a priori} evaluation of the existence of 
these components is required. As we take the approach that bulges in S0s
can be either classical bulges with the $R^{1/4}$-law type profiles or 
pseudo-bulges, the concept of a bulge is not self-evident. For example 
there are views according to which pseudo-bulges are actually evolved 
bars seen edge-on (Athanassoula 2002, 2005): when the bar is created it 
is thin, and when it evolves in time its vertical extent increases
particularly in the inner part of the bar and the morphology turns to 
a boxy or peanut-shaped structure. It has also been shown by Samland $\&$
Gerhard (2003) that bulges formed at different times may appear 
in the same galaxy. 

In spite of this phenomenological problem, some rules of thumb can be used 
to identify the different components. We particularly follow KK04, who
define a {\it pseudo-bulge} as a nearly exponential structure ($n$ = 1-2), 
which in some cases might have a boxy/peanut shaped structure and 
particularly, 
its flattening is similar to that of the outer disk. {\it Lenses} (or ovals 
in spirals) might have similar ellipticities as pseudo-bulges (typically 
b/a $>$ 0.85), but in distinction to pseudo-bulges, they generally have 
lower surface brightnesses and rather sharp outer edges (KK04). 
{\it Inner disks} can be clearly identified, if the galaxy has near-nuclear 
spiral arms. In unclear cases the elliptical inner structure is assumed 
to be an inner disk if its position angle $\phi$ differs from $\phi$ of 
the outer disk by less than $10^0$ (see Erwin $\&$ Sparke 2003, 
hereafter ES03).    
In order to identify the different components we measure {\it (1) the 
radial profiles of the ellipticities and position angles}, {\it (2) the 
radial profiles of the low azimuthal wavenumber Fourier amplitudes and 
phases}, and construct (3) {\it unsharp masks}. Bars, lenses and inner disks
all appear as bumps in the ellipticity profiles, if their surface 
brightnesses are high compared to that of the underlying disk. The Fourier 
method is sensitive for detecting weak bars and ovals, whereas the unsharp 
masks are capable of showing also the innermost structures of galaxies.

\subsection{The orientation parameters}

$B$-band images were used to measure the radial profiles of the 
isophotal major-axis position angles ($\phi$) and the 
minor-to-major axis ratios 
($q$ = b/a). The position angles and inclinations of the disk were 
estimated from the mean values in the outer parts of the disks. If no 
$B$-band images were available the orientation parameters were estimated 
using $K_s$-band images: these images were first compared with the 
Digitized Sky Survey plates to ensure that 
the outer parts of the disks are visible. The surface brightness profiles 
and the orientation parameters were derived using the ELLIPSE routine 
in IRAF (Jedrzejewski 1987): a linear radial scaling in steps of one 
pixel was used to have good spatial resolution particularly in the outer 
parts of the disks.
In order to minimize the effects of noise and contamination by bad pixels 
and cosmic rays, deviant pixels above 3 $\sigma$ were rejected. The measured 
disk position angles and axial ratios are shown in 
Table 2, where the uncertainties are
standard deviations of the mean calculated in the radial range indicated 
in the table. The outer disk of NGC 2781 is so weak that the ELLIPSE 
routine failed. In this case the orientation 
parameters were determined manually by adjusting 
an ellipse to the outer isophotes.
For comparison, the table also shows the 
orientation parameters given in RC3 and those obtained by ES03. 
The $q$ and $\phi$ profiles are 
shown in Fig. 5 with a logarithmic radial scale, in order to better 
illustrate the presence of the different structural components.

The orientation parameters found in this study are generally in good 
agreement with those obtained by ES03, who used high resolution $R$-band 
images for their measurements. We have 9 galaxies in common with their 
sample and good agreement was found for 7 of the galaxies.
However, for NGC 2859 we measure a considerably smaller $q$-value 
(0.76 vs. 0.90): it seems that the image used by ES03 ends up to the 
outer ring, while we measure also the extended disk outside the ring 
thus giving a more reliable estimation of $q$. Also, the position angles 
of NGC 1022 and NGC 2681 are completely different in the two studies, 
but this is not very surprising taking into account that these galaxies 
have almost circular outermost isophotes.
While comparing the measured orientation parameters with those given in RC3 
larger deviations were found, which is expected because RC3 orientation
parameters are based on photographic plates, which do not have the same 
depth and quality as the modern CCD images.

\subsection{Fourier decomposition and unsharp masking}

Fourier decompositions are calculated from the $K_s$-band images in different 
radial zones and the amplitude and phase of each component is tabulated 
as a function of radius (Salo et al. 1999; LSBV04). 
The Fourier modes up to m = 10 were calculated, although the main modes 
in bars and ovals are m = 2 and m = 4. Bars are identified mainly by assuming 
that the phases of the m = 2 and m = 4 amplitudes are maintained 
nearly constant
in the bar region, in distinction to spiral arms, where the phase changes 
as a function of radius. Another useful way of identifying bars is to 
map the galaxy image into a log polar coordinate system:
a bar appears in the image either as a linear structure or as a bright spot,
depending on the radial surface brightness profile of the bar. 
A contrast between the bar and the surrounding region can be highlighted 
by subtracting the m = 0 component from the image. The polar angle maps 
as well as the amplitudes 
and phases of the m = 2 and m = 4 Fourier modes are shown for all 24 galaxies 
in Fig. 1. Before calculating the Fourier modes, the images were deprojected
to face-on orientation using the method described in LSBV04:
we use a 2D decomposition method to estimate
the relative flux of the bulge component, which is subtracted from 
the original image. In these decompositions a bulge model with 
circular isophotes is used.
The image is then deprojected to face-on orientation after which the 
bulge is added back by assuming that it has a spherical 3D light distribution.
This deprojection method was preferred in order to avoid artificial
stretching of nearly spherical bulges while deprojecting 2-dimensional images.
As bulges in some galaxies are not circular, this might also
be the reason for the large $A_m/A_0$-ratio at the very center in
some of the galaxies in Fig.1. Notice that in the decompositions shown 
in Section 4 the bulges are not restricted to have circular isophotes.

An unsharp mask works as a filter suppressing large-scale low-frequency 
variations in the images (Malin $\&$ Zealey 1979; ES03). In this study a 
mask was created by making a smoothed copy of the original $K_s$-band image, 
which was then subtracted from the original image. Typical windows used 
for smoothing the images varied between 5 to 20 pixels, depending on the 
size of the inner structure of the galaxy. In principle it is also 
possible to divide the original image with the smoothed image, but that 
was only occasionally done. Due to increased contrast between the 
non-axisymmetric structure and the surrounding region, it is possible 
to better estimate the inner morphology of the galaxies, for example by 
distinguishing secondary bars from inner rings and inner spiral arms, 
which all appear in a similar manner in the $q$-profile. 
The unsharp masks are shown in Fig. 5.

\subsection{Discussion of individual galaxies} 

The identified structural components are listed in Table 3, where the 
primary bars are denoted as $bar_1$, the secondary bars as $bar_2$, and 
the tertiary bars as $bar_3$, and if only one bar appeared in a galaxy, 
simply as $bar$. In case of three bars, as in NGC 2681, $bar_2$ is the 
most prominent bar in the galaxy and $bar_3$
denotes the extremely faint bar at a larger radius.
Following KK04, if a galaxy has a morphological type of 
S0 - S0/a the flat inner structure is called a {\it lens} and for later 
morphological types it is called an {\it oval}. The number in parentheses 
indicates the semi-major axis length of the structure in arcseconds. This 
length has been estimated from the phases of the m=2 Fourier modes
which give systematically slightly longer bars than estimated from the minima
in the $q$-profiles (Laurikainen $\&$ Salo 2002).
Below we discuss observational evidence for these components for 
most galaxies. As discussed above, the original images appear 
in Fig. 4 (upper left panel), the Fourier amplitudes and phases
in Fig. 1, and the unsharp masks and 
the radial profiles of the orientation parameters in Fig. 5.

{\bf NGC 718}: This galaxy has a primary bar extending to $r$ = 20 arcsec 
and also shows weak evidence of a secondary bar at $r$ = 5 arcsec 
(Figs. 1 and 5), 
previously detected in the $R$-band by ES03 as an elliptical inner component.  
Both structures appear as minima in the $q$-profile and in the case of 
the primary bar, also as a rapid change in the position
angle at the end of the bar. Both features have significant peaks in 
the m = 2 and m = 4 amplitudes of density. For the primary bar the m = 2 
and m = 4 phases are also maintained nearly constant in the bar region, 
and in the polar angle map the bar is identified as an intensity maximum 
at $r$ = 20 arcsec. The weak innermost structure can be detected also
in the polar angle map and marginally in the unsharp mask.

{\bf NGC 936}: Characteristic for this galaxy is a prominent bar, 
an oval and a small nearly spherical bulge. In the central part of the galaxy
there is also a small elliptical structure (Fig. 5), which was identified as 
a nuclear ring by ES03 in a high resolution (0.11 arcsec) $R$-band image.  
The bar and the inner elliptical structure are both identified in the
Fourier analysis, and the primary bar also as a minimum in the radial  
$q$-profile.

{\bf NGC 1022}: NGC 1022 is a peculiar dusty ringed galaxy  
discussed also by ES03. It is classified as a barred galaxy both in RC3 and by 
ES03, but our Fourier analysis shows that the phase of the m = 2 amplitude is
not maintained constant in the assumed bar region, indicating
a spiral-like nature of this structure, seen also in the 
polar angle map. Other similar cases among spiral
galaxies have been earlier discussed by LSBV04.
This galaxy has also a slightly
flattened oval inside the bar, and a pseudo-ring surrounding the 
bar-like structure.  


{\bf NGC 1415}: This galaxy is classified as having an
intermediate type bar in RC3, and it is also reported to have an elliptical 
inner structure
in the red continuum by Garcia-Barreto $\&$ Moreno (2000) and in the 2MASS 
images by Erwin (2004). Garcia-Barreto $\&$ Moreno interpreted it
as a secondary bar, but due to the similarity in position angle with the
main disk, Erwin suggested that it might be an inner disk. 
We found near nuclear spiral arms at $r$ = 9 arcsec, being well illustrated 
in the unsharp mask (Fig. 5) and visible also in the polar angle map, 
which confirms the disk-like nature of this structure. The primary bar
has ansae at the two ends of the assumed bar, but the morphology 
has some disk-like characteristics. 
Our $K_s$-band image is not very deep, showing only the bar region but  
not the outer exponential disk.

{\bf NGC 1440}: This galaxy has a classical bar detected by all our criteria.
There is also a large lens inside the bar and a small, almost spherical bulge.

{\bf NGC 1452 }: NGC 1452 is a barred galaxy having a prominent ring around 
the ends of the bar, a large oval inside the bar, and a small bulge.

{\bf NGC 2196}: This is a non-barred galaxy, but has an elliptical
inner structure, which is not completely aligned with the underlying disk.
The inner elliptical structure appears also as a strong m = 2 peak in the amplitude 
profile in the Fourier analysis.

{\bf NGC 2273}: This is a famous barred galaxy with four rings 
(see e.g. discussion
in Buta $\&$ Combes 1996), studied previously 
by ES03 in the optical and using NICMOS HST images. They showed that, besides 
the bar, this galaxy has also nuclear spiral arms inside a nuclear ring. 
We confirm the presence of the near nuclear spirals in our ground-based 
infrared image at $r$ = 2-3 arcsec (Fig. 5), and these arms are clearly 
visible also 
in the polar angle map. The bar is well detectable in the polar angle map, 
but the m = 2 phase is maintained nearly constant at a much larger radius 
from the galaxy center than the bar region alone, mainly because
this galaxy has rather open spiral arms outside the bar.  
The inner disk is surrounded by an oval or a flattened bulge.

{\bf NGC 2460}: This galaxy is classified as a non-barred galaxy in RC3, 
but clearly 
has an elongated inner structure at $r < $10 arcsec, detected as a bar-like 
structure in the $q$ and $\phi$-profiles and in the Fourier analysis. 
As the position angle of this structure deviates even by $25^0$ from that 
of the main disk, it might be a weak secondary bar. 

{\bf NGC 2681}: NGC 2681 has been found to be a triple barred galaxy by ES03 and by 
Erwin $\&$ Sparke (1999), based on both ground-based and HST NICMOS images. 
All three bars are detected also in this study, as minima in the $q$-profile 
and as blobs in the polar angle map.
This galaxy also has a nearly round lens at the radius of the secondary bar.

{\bf NGC 2781}: NGC 2781 is classified as a weakly barred galaxy 
in RC3. Characteristic for this galaxy is an extremely faint outer disk  
and elliptical structures at the distances of $r < $ 45 arcsec and 
$r < $ 10 arcsec from the galaxy center.  
Both elliptical structures are visible as minima in the $q$-profile 
and as intensity maxima in the Fourier analysis. The unsharp 
mask (Fig. 5) shows that the smaller structure is actually an inner 
disk with a two-armed spiral.
The outer elliptical is most probably a bar, but does not have 
the typical rectangular shape of a classical bar, usually associated
with a small axial ratio (in this case $q$ = 0.52).
Both the inner and outer elliptical structures deviate 15$^0$
from the orientation of the outer disk.


{\bf NGC 2859}: In the optical this galaxy is classified as a double barred 
galaxy (Kormendy 1979; Wozniak et al. 1995), 
having also inner and outer rings (RC3). We confirm the double barred nature 
in the near-IR, which is obvious using all our criteria for bars.
In the unsharp mask the secondary bar has a rectangular morphology typical
for a classical bar. This galaxy has also two
ovals, one at the radius of the primary bar, and another inside that bar.

{\bf NGC 2911}: NGC 2911 clearly has no bar/oval components, only a disk and a 
slightly flattened bulge. However, it has a peculiar inner structure
with a tiny polar edge-on disk in the very center (Sil'chenko $\&$ Afanasiev
2004).

{\bf NGC 2983}: This galaxy has a strong bar, a central elliptical 
structure, and a lens
inside the bar. The primary bar is well visible in the $q$-profile
and in the Fourier analysis, whereas the inner elliptical appears
only as a density peak in the m = 2 amplitude profile. The secondary
bar can be identified also in the unsharp mask (Fig. 5). 
The primary bar has an ansae-type morphology
with blobs at the two ends of the bar.

{\bf NGC 3414}: This is a peculiar galaxy, which looks like a barred system
with a large oval, but different interpretations have been given of its
true nature. For example, Whitmore et al. (1990) suggested that it is a
galaxy seen edge-on with a large-scale polar ring in the $R$-band, whereas 
according to BBA98 and Chitre $\&$ Jog (2002) it might be a nearly face-on galaxy
with a prominent bar.   
 
{\bf NGC 3626}: In RC3 this galaxy is classified as a non-barred system 
with an outer ring. However, both the $q$-profile and the Fourier analysis 
show the presence of a bar at $r < $ 45 arcsec, having a morphology with ansae 
at the two ends of the bar (Fig. 5). This galaxy has also an inner elliptical,
detected as a bar-like structure at $r$ = 5 arcsec by the Fourier analysis, 
and as an elliptical feature in the unsharp mask. The position angle
of this structure deviates from that of the main disk by 10$^o$, and might
actually be an inner disk.

{\bf NGC 3941}: This galaxy is classified as a double barred system by ES03 in the optical 
region. In the $K_s$-image the main bar has ansae-type morphology and all 
characteristics of a bar. However, the unsharp mask shows that the inner 
elliptical is rather an inner disk showing two-armed spirals (Fig. 5). 
Both the inner disk and the bar appear as minima in the $q$-profile
and are identified as intensity peaks in the Fourier analysis. This 
galaxy has also an oval inside the bar. 

{\bf NGC 4245}: In RC3 this galaxy is classified as a barred galaxy with an inner ring. 
ES03 found a nuclear ring in $R$-band, but no evidence of a secondary bar in 
the optical HST image. ES03 also report dust lanes
in the inner ring leading into a nuclear spiral that continues into 
the nucleus. The nuclear ring at $r$ = 5 arcsec is visible also in our unsharp 
mask of the $K_s$-band image (Fig. 5). 
NGC 4245 has an oval surrounding the secondary bar.

{\bf NGC 4340}: In the optical region this galaxy has been classified as a 
double barred system by Erwin (2004), the secondary bar being surrounded by a
nuclear ring. Both bars are identified also in our $K_s$-band image,  
by all indicators of a bar. The secondary bar is aligned with the primary bar,
which deviates from that of the underlying disk.
This galaxy has also a nearly spherical bulge and an oval inside the 
primary bar. 


{\bf NGC 4608}: This galaxy is classified as a barred system with 
an inner ring in RC3. Both components are visible also in our $K_s$-band 
image. Additionally, NGC 4608 
has a large oval inside the bar, and apparently a small spherical bulge. This 
galaxy has been reported as a candidate of barred
galaxies without any sign of the disk component (GS03) and will be 
discussed in detail in Section 4.  

{\bf NGC 4643}: This is a famous barred Polar Ring galaxy (Whitmore et al. 1990),  
studied previously also by ES03. They found an elliptical
inner structure in the optical region, and based on an unpublished high 
resolution $H$-band 
image by Knapen that structure was interpreted to be an inner ring.
Our $K_s$-band image shows four knots in the central part of the galaxy, 
which most probably indicates the presence of a nuclear ring (Fig. 5).

\section{Multicomponent 2D-decompositions}

For structural decomposition of galaxies, three main types of methods
have been generally used: (1) one-dimensional (1D) methods, where
either a major axis surface brightness cut or an azimuthally averaged
profile is fitted by assuming various functions for the bulge and the
disk, (2) the Kent (1985) method where the bulge and the disk are
separated based on their different isophotal ellipticities, and (3)
two-dimensional (2D) methods, where the model functions for the bulge
and the disk are fitted simultaneously to all pixels of the image.
The advantage of using azimuthally averaged profiles and a 1D method is
the small amount of noise in the profile, whereas Kent's method is
powerful in separating the bulge without any specific assumptions of
the functional form of the bulge or disk. However, for galaxies with
strong non-axisymmetric structures like bars both methods might fail,
because omitting a large bar in the decomposition might
overestimate the light attributed to the bulge model
(LSBV04; see also discussion in Section 4.2). 

The first 2D methods used the de Vaucouleurs's $R^{1/4}$ law for the
bulge (Byun $\&$ Freeman 1995; Shaw $\&$ Gilmore 1989; de Jong 1996;
Wadadekar, Robbason $\&$ Kembavi 1999), but later
studies have shown that the more general S\'ersic's $R^{1/n}$ function
(Sersic 1968) can better account for the bulge profiles. This has
been shown for the bulges of late to intermediate type spirals
(M\"ollenhoff $\&$ Heidt 2001; Simard et al. 2002; MacArthur, 
Courteau $\&$ Holtzman 2003), and also for the
bulges of S0s (D'Onofrio, Capaccioli $\&$ Caon 1994; 
de Souza, Gadotti $\&$ dos Anjos 2004) and even for the surface
brightness profiles of elliptical galaxies (Caon, Capaccioli $\&$ D'Onofrio 
1993).

In spite of the fact that the 2D method is a powerful tool for
separating non-axisymmetric structures in galaxies, that advantage has
been only occasionally used
(de Jong 1996; Prieto et al. 2001; Peng et al. 2002; LSBV04). 
In the following our
method together with some tests will be briefly described, and its
application for the galaxies in our sample are discussed.

\subsection{The algorithm}

We use a 2D decomposition method where, in addition to a bulge and a disk,
up to three non-axisymmetric structures can be simultaneously
fitted. For the radial profile of the bulge we use a S\'ersic function,
while the disk is always assumed to be exponential; for the bar-like
structures either a Ferrers or a S\'ersic function is used. 
No inner truncation is used for these components. An
exponential inner disk and a nuclear Gaussian point source can also be
included, so that the maximal number of fitted components is seven
(the maximal number of free parameters is 31). Due to practical
reasons related to computing time and difficulties caused by the
possible degeneracy between the fitted components, this seems like a
reasonable upper limit. Also, the method is flexible in the sense that
the user can decide which of the model parameters are fixed and which
are free variables in the fit; also any number of the components can
be omitted from the fit.

The fitting is performed in flux-units, which makes it possible to take
into account also the faint outer regions of the images, where some
fraction of the pixels have negative values. The best solution is
found iteratively, where fitting to the data is accomplished by
minimizing the weighted squared deviations of the data from the fit:



$$
\chi^2 \ = \sum_{i}^{N_{pix}} w_i (F_{i} - F_{model} )^2,\eqno\stepeq
$$

\noindent where $F_i$ and $F_{model}$ denote the measured and modeled
surface brightnesses at each image pixel, $i=1, ..., N_{pix}$, and
$w_i$ is the weight assigned to each pixel. The minimization of the
deviations between the original image and the model is made with the IDL
procedure CURVEFIT.

Depending on the choice of the weighting function, more weight can be
assigned to the pixels in the central parts of the galaxy where the
surface brightness is high, or it can be used to emphasize the faint
outer disk. One natural choice would be to use a weight related to the
distribution of noise in the image: $w_i$ = $1 / {\sigma_i}^2$, where
$\sigma_i$ denotes the standard deviation of $F_i$. In the case of
Poisson noise this would imply $w_i$ = $1/F_i$. The effects of the
weighting function in the decompositions will be discussed in the next 
section. In order to
account for the effects of $seeing$, the model is convolved with a
Gaussian PSF using a FWHM measured for each observed image. No attempt
was made to correct the effects of dust, but the problem is not
very serious in the near-IR, where the extinction is minimal, and
particularly it is not serious for early-type galaxies that generally
have only a small amount of dust.

All components (including the bulge) are allowed to follow a generalized 
elliptical shape (Athanassoula et al. 1990), defined by the equation:

$$
r=(|x|^{c+2} + |y/q|^{c+2})^{1/(c+2)}.\eqno\stepeq
$$

\noindent The shape of the isophote corresponding to $r$ = constant is
boxy when the shape parameter $c > $ 0, disky when $c < $ 0, and purely
elliptical when $c$ = 0; circular isophotes correspond to $c$ = 0 and an
axial ratio $q$ = 1. Here $x$ and $y$ are the rectangular coordinates in
a system aligned with the major axis of the component in question,
defined by its position angle $\phi$.

A S\'ersic's function is used to describe the brightness profile of
the bulge:
$$  
I_b(r_b) \ = \ I_{0b} \exp[-(r_b/h_b)^{\beta}],\eqno\stepeq
$$

\noindent where $I_{0b}$ is the central surface density, $h_b$ is the
scale parameter of the bulge, and $\beta$ = $1/n$ determines the slope
of the projected surface brightness distribution of the bulge. The
coordinate $r_b$ is the isophotal radius defined by
Eq. 2, using the parameters $q_b$, $c_b$ and
$\phi_b$, where the subscript stands for the bulge. In the case of
an elliptical bulge, $\phi_b$ is its major axis position angle measured
counter clockwise from North in the sky plane. Special cases of S\'ersic's
formula are the exponential function with $n$ = 1 and the de
Vaucouleurs $r^{1/4}$ law with $n$ = 4. Note that since the numerical
value of $h_b$ depends strongly on $\beta$, we typically describe
our fitted bulge models with the bulge effective radius $r_{eff}$
(radius of the isophote that encompasses half of the total bulge light).

Bars (and ovals/lenses) are fitted using a function, which has the
form:
$$  
I_{bar}(r_{bar}) \ = \ I_{0bar} \ (1-(r_{bar}/a_{bar})^2)^{n_{bar}+0.5}, \ \ r_{bar}<a_{bar} 
$$
$$
\hskip 1.5cm    =  0, \ \ \ r_{bar}>a_{bar},\eqno\stepeq
$$

\noindent where $I_{0bar}$ is the central surface brightness of the
bar, $a_{bar}$ is the bar major axis, and $n_{bar}$ is the exponent
of the bar model defining the shape of the bar radial profile.  The
isophotal radius ($r_{bar}) $ is defined via parameters $q_{bar}$, $c_{bar}$,
$\phi_{bar}$ in Eq. 2.  For bar components we have
two choices for the reference plane: besides the sky plane
one can also specify the bar shape and orientation parameters using
the disk plane as a reference plane, in which case $\phi_{bar}$ is
counted counter clockwise along the disk plane from its nodal line.
The function (Eq. 4) corresponds to a projected surface density
of a three dimensional prolate Ferrers bar, with $a > b$ = $c$, seen
along the $c$-axis. Thus the possible 3D structure of a bar is not
taken into account. Alternatively, the radial profiles of bars 
(and ovals/lenses) can be described by a S\'ersic's function. 
The advantage of
using a projected Ferrers function is that it is rather flat and that
the surface brightness drops near the outer edge, as is the case also
for bars and ovals in real galaxies. Taking into account that bars
typically have rectangular shapes, more realistic bar models are
obtained if the shape parameter $c_{bar}$ is also added. The Ferrers function is
expected to work well especially for bars in early-type galaxies,
which are known to have flat surface brightness profiles
(Elmegreen $\&$ Elmegreen 1985). But for some galaxies a S\'ersic's 
function with a more centrally peaked profile might also give reasonable fits.
Disk components are described with an exponential function
$$ 
I_d(r) \ = \ I_{0d}  \exp[-(r/h_r)],\eqno\stepeq
$$

\noindent where $I_{0d}$ is the central surface density of the disk,
and $h_r$ is the radial scalelength of the disk. For the disk the radius $r$
is always calculated along the disk plane, defined by the position
angle $\phi_d$ and axial ratio of the disk $q_d$. In principle,
$\phi_d$ and $q_d$ can also be free variables in the fit; however, in
the current study they are fixed to the values derived from
observations like those described in section 3.1. Inner disks
can be fitted by an exponential function, and as for
bars the vertical thickness of the disk component is ignored. In principle,
active nuclei and other bright central sources can be fitted 
with a Gaussian PSF, but that is not done in this study. 

A three-component version (bulge/disk/bar) of this algorithm has been
previously used in the studies by
LSBV04, Laurikainen, Salo $\&$ Buta (2004) and by 
Buta, Laurikainen $\&$ Salo (2004).

\subsection{Testing the method}

The decomposition method was tested by applying it to synthetic
images, created using a S\'ersic's function for the bulge, and an
exponential function for the disk, and in some cases also a Ferrers'
function for the bar. In order to have realistic images Poisson noise
and background noise were added, and the images were convolved with a
Gaussian PSF, $\sigma_{PSF}$ (=FWHM/2.355) imitating the typical 
$seeing$ in the science images. A
typical inclination of the disk ($30^o$) was also assumed. If not
otherwise mentioned the resolution in the test images was taken to be
similar to that in the original science images. Decompositions for the
test images were performed starting from initial parameter values that
deviated significantly from the true parameters of the created images.

{\it The effect of the weighting function} on the scalelength of the
disk was investigated for a synthetic galaxy image with $\beta$ = 0.5
($n$ = 2) and bulge-to-disk ratio $B/D$ = 0.2. The bulge effective radius
$r_{eff}$ = 3 and the disk scalelength $h_r$ = 30 pixels. The amount of
noise was specified by setting the signal-to-noise ratio $S/N$ = 3.0 at
the distance of 4$h_r$ (separately for the Poisson and the background
noise component, so that the total $S/N$ is in fact $\sqrt{2}$ times
smaller). Another case with twice larger noise level ($S/N$ = 1.5) was
also examined, representing an already very noisy image. A $seeing$ with
$\sigma_{PSF}$ = 1 was added.
The results for different weighting functions are given for
two different $S/N$-ratio in
Tables 4a and 4b, using a large range of possible $r$ and $F$
combinations, where $r$ denotes the pixel distance from the center
along the disk plane and $F$ is the corresponding flux: note the
difference between $F$ and $F_{model}$ in the table, the former being the
observed pixel value while the latter is the corresponding model value
for this pixel. In the case the pixel flux $F$ was below a 3$\sigma$
noise level it was replaced with this value when used in a weighting
function. The rows in this table are ordered according to an
increasing relative weight given to the outer parts of the galaxy
(i.e. $w$ = $1/r^2$ weights very strongly the inner galaxy while $w$ =
$r/F_{model}$ places more weight on the outer disk).

We found that the decomposition works
well in this simple bulge/disk case, regardless of the exact choice
of the weighting function. 
For the images with high S/N-ratio poor results are obtained only for 
the function
$w$ = $1/r^2$, giving more weight to the inner pixels:
it tends to underestimate the scalelength of the disk
and gives large deviations from the true shape parameters of the bulge.
Besides $h_r$, these weighting functions also give
nearly correct effective radii for the bulge component, within 
an accuracy of 0.2$\%$. Also, the $B/D$-ratio is independent 
of the weighting function used. 
For noisy images poor results are generally obtained when weighting 
functions based on the observed $F$ were used. But even in this case 
the results are fairly accurate for weighting functions like 
$w$ = $1/(F_{model}r)$, $w$ = $r$, or
$w$ = $1/F_{model}$. In this case the results with
$w$ = $r/(F_{model})$ also become sensitive to the starting values of the
iteration, indicated by the empty bin in the table. The behavior of
the weighting functions involving the observed fluxes $F$ is also
fairly interesting: particularly in the case of larger noise it is
evident that fits with these weighting functions tend to converge to
values of $h_r$ which are {\it systematically} too low. 
The above tests show 
that neither the $B/T$-ratio nor the parameter $\beta$ of the bulge 
are sensitive to the weighting function used: the decomposition 
becomes sensitive to the weighting function only if the image is
very noisy, or as will be discussed later, if complicated galaxy morphology
is fitted by a simple decomposition model.


Next {\it the effect of seeing on the parameters of the bulge} was
investigated. We first checked that the original parameters of the
synthetic images, smeared by $seeing$, are very accurately recovered
when the correct $\sigma_{PSF}$ is used in convolving the model 
functions. On the
other hand, if the $seeing$ correction is not included in the fit, or an
inaccurate $seeing$ correction is applied, substantial errors in the
derived bulge parameters are possible. Table 5 collects some results,
using synthetic images that contained only a bulge component, with
$\beta$ = 0.25, 0.50, and 1.0, thus ranging from de Vaucouleurs's type
bulge ($n$ = 4) to an exponential bulge with $n$ = 1. No noise is added, and
in order to assure that the pixel resolution does not cause problems,
a large effective radius of $r_{eff}$ = 20 pixels is used. According to
this table, the case of a poor $seeing$ combined with a compact bulge
can lead to a serious under or overestimate of $\beta$ if the $seeing$ in 
the image is not well determined. If no $seeing$ correction is made the effect
of $seeing$ is to make the bulge more exponential than it actually is:
as an extreme case, a de Vaucouleurs type profile ($n$ = 4) appears as an
exponential bulge ($n$ = 1).

As one of the key issues in this study is to apply multicomponent decomposition
to galaxies with prominent bars and ovals, we also checked {\it how much 
omission of the bar in the
decomposition may affect the $B/T$-ratio}. Synthetic test images were 
created (1) having only
a bulge and a disk, and (2) then adding a large bar to that galaxy model. 
A Ferrers function for the bar was used, the bulge parameters were taken to 
be $\beta$ = 0.60 and $h_b$ = 1.0, and the scalelength of the disk $h_r$ = 45. 
Again, in order to create realistic images, $seeing$ and noise 
were added to the synthetic data, comparable to those used in the tests 
in Table 4.
The decompositions for these two synthetic images are shown in Fig. 2. 

As expected, applying a bulge/disk decomposition
to the first test image recovers the bulge and the disk parameters with
relatively high accuracy (Fig. 2a). Particularly, we found
$B/D$ = 0.38, in comparison to the input $B/D$ = 0.37 in the synthetic 
image.  For
the second test image two types of decompositions were made, namely
bulge/disk (Fig. 2b) and bulge/disk/bar decompositions (Fig. 2c), both
giving relatively good looking fits for the surface brightness profile. 
We found
that when fitting all three components, the correct $B/D$ = 0.37 was
returned. However, if only a bulge and a disk were fitted, $B/D$ = 0.57
was obtained, which overestimates the $B/D$-ratio even by 1/3.  
It is evident that in the bulge/disk decomposition a significant fraction 
of the bar flux was erroneously assigned to the bulge, 
illustrated in the residual image where both positive 
and negative pixel values are visible. The negative residuals appear 
due to the fact that too much light is subtracted in a region slightly 
exceeding the radius of the bar, whereas the positive 
residuals appear because the bar itself was only partially subtracted.

As the above galaxy model has a small $B/D$-ratio with a small spherical
bulge, other models with higher $B/D$-ratios were also studied (see Table 6).
Three galaxy models were created with $B/D$-ratio near unity: in {\it model 1} 
the bulge was taken to have $\beta$ = 0.5, whereas in {\it model 3} 
$\beta$ = 0.33. One of the models ({\it model 2})
also tested the effect of a non-spherical shape of the bulge 
($q$ = 0.9) in the 
decomposition. All galaxy models also included a prominent bar modeled by a 
Ferrers function with $n_{bar}$ = 2 and an ellipticity $q$ = 0.3. This 
bar contributed to the total flux by (bar light)/(total light) = 0.1.
As above, also in this case $seeing$ and a small amount 
of noise were added to the images. The created images were fitted either
by bulge/disk or bulge/disk/bar decomposition. In order to simultaneously 
test also the effect of the weighting function, all test images were 
decomposed with three different weighting functions: $w_i$ proportional
to $1/F_i$ or $r_i$, or the same weight was used for all pixels.
We found that when all three components were fitted simultaneously the true 
$B/D$ and $\beta$-values were recovered with a high accuracy for 
all three models. 
The solutions were also practically independent of the weighting 
function used. However, if only a bulge and a disk were fitted  
large deviations from the true $B/D$ and $\beta$-values appeared 
(the deviations from the true values are indicated in parenthesis).
In this case also the weighting function started to be important:
for example using a constant weight for all pixels in the image, {\it model 3} 
gave a solution where the whole profile could be fitted nearly by one single 
function. While comparing the two models which deviated only by the 
ellipticity of the bulge ({\it models 1} and {\it 2}), similar 
results were obtained.
Based on the above tests it seems that by omitting a prominent
bar in the decomposition systematically overestimates the $B/T$-ratio,
assigning a significant fraction of the bar flux to the flux
of the bulge. 

S0 galaxies generally have weak disks and therefore it is important to
study {\it the limiting cases where the disk might be erroneously
lost} in the decomposition, for example because they are overshadowed 
by luminous bulges or
because the disks are too noisy. An example of a galaxy model with a
massive rather exponential-like bulge, but having a nearly
$R^{1/4}$ law type total surface brightness profile is shown Fig. 3a:
in this model the bulge has $\beta$ = 0.5 (or $n$ = 2.0), the disk has
$h_r$ = 50, and $B/D$ = 1.4. Also, the image is truncated already at 100
pixels, or two disk radial scalelengths. In spite of the lack of a
visually detectable exponential part in the surface brightness profile
our method still finds the scalelength of the disk with a high
accuracy ($h_r$ = 51). If a similar profile is created, but using
$\beta$=0.3 (or $n$ = 3.3) and higher noise level is added, it is
possible to miss the disk, but only if completely erroneous initial
parameters are given for the bulge (Fig. 3b). In this case the disk is
lost suddenly after several iterations, but the final model is far
from the surface brightness profile of the created image, showing that
a single function is not capable of accounting for this profile.
Finally, Figs. 3c-d show tests using a galaxy model having a weak and
noisy disk dominating the outer part of the galaxy, and a rather small
bulge with a nearly de Vaucouleurs's type surface brightness profile
($\beta$ = 0.3). The two decompositions shown for this model are
otherwise similar, except that different weighting functions are
used. We found that while it is not possible to miss the disk in these
decompositions, a bad choice of the weighting function can lead to
considerable errors in the fitted disk scalelength: in Fig. 3d we
used the instrumental Poisson weighting $w$ = $1/|F|$, without
any lower noise limit (the absolute value was taken to avoid negative
weights), whereas in Fig. 3c $w$ = $r$. 
Incidentally, these test images resemble the galaxy NGC 2911 in our sample.
It appeared that the disk was properly found using $w$ = $r$, but not 
when the weight is proportional to $1/F$. However, $w$ = $1/F_{model}$ recovers
the correct parameters with the same accuracy as the weight $w$ = $r$.  

\subsection{The fitting procedure}

In principle the decompositions can be performed by fitting all 31 variables
simultaneously, but in practice this is not reasonable or even possible. 
Instead we proceed in steps where the 
orientation parameters are first fixed to the values obtained by 
fitting ellipses to the isophotes, as described in Section 3.1. Also, 
in order to have control of the physical meaning of the different 
components, {\it a priori} identification
of the structural components is made while making the initial guesses for the 
fitted functions. If bars and ovals appear 
in the same galaxy, an attempt was made to avoid possible degeneracy 
between these components by using different types of radial profiles 
for bars and ovals. In some galaxies bars
might have a thin rectangular component together with a thicker and 
shorter component, in which case it is possible to use two different 
functions for the bar. However, in this study it is not critical whether 
the thick component is part of the bar, or an oval inside the bar.
 
Our images are generally deep enough for reliable estimation of the 
scalelengths of the disk, but due to the 2D nature of the method the 
surface brightness profiles are often rather noisy. In principle a lower 
noise level could be achieved by rebinning the images, but in that case 
the advantage of the high image resolution is lost. Partly for this reason 
and also for saving time, the decompositions
were performed in steps in the following manner: 

(1) Generally the scalelength of the disk, $h_r$, was measured by doing 
bulge/disk decomposition for the image rebinned by a factor of 4, 
but if a prominent bar appeared in the galaxy, the bar was also fitted. 
$Seeing$ was taken into account and a weighting function was selected to
fit well particularly the exponential disk. Generally the pixels were
weighted inversely proportional to Poisson noise (to $1/|F_i|)$),
but in some cases $w_i$ proportional to $r_i$ was more useful.
In most cases the images were deep enough for reliable estimation of $h_r$, 
but for the galaxies NGC 1415, NGC 2911, and NGC 4340, the obtained $h_r$ 
is most probably underestimated.

(2) As a next step $h_r$ was fixed and multicomponent 
decomposition was performed. In order to have a good fit in a reasonably 
short time the shape parameter describing deviations from elliptical 
isophotes was not used. Again, a Gaussian function was used in the 
decomposition to correct for $seeing$, and this time a weighting function 
was always taken to be inversely proportional to Poisson noise.  

(3) Finally a decomposition including also a non-zero shape parameter 
$c$ of the bulge 
and allowing for apparent bulge ellipticity was performed, 
using the previous solution for the initial parameters.  
The model functions were then subtracted from the original 
image and the residual image was inspected. If the residuals were too large the
procedure was repeated a couple of times until a satisfactory solution
was found. Taking into account the large 
number of free parameters in the fit it was sometimes useful 
to repeat the decomposition several times fixing some of the parameters 
at each step.

\subsection{The decompositions for individual galaxies}

The structural decompositions derived for the $K_s$-band images are 
shown for all galaxies in our sample in Fig. 4. We display 
the original and model images (upper and lower left panels, respectively), 
and the surface brightness profiles together with
the model functions (right panel). Also, the residual images obtained 
by subtracting the model functions from the original images
are shown in Fig. 5 (left panel).
The most important parameters of the best fitting solutions are 
shown in Table 7.
In the table the columns ``Ferr 1'', ``Ferr 2'' and ``Ferr 3'' show
the parameters of the Ferrers function, corresponding either to bars
or ovals, depending on the physical components appearing in each galaxy.
Bulge-to-disk light ratios in the table are given so that only
the exponential component is included in the disk. However, it would be 
more realistic to use $B/D_{tot}$,
where bars and ovals/lenses are also included to the disk 
(which can be calculated from the parameters given in the table).   
For NGC 3414 no decompositions were made, because it was not clear whether 
this is a face-on or edge-on galaxy. In general
we take a conservative approach in a sense that if the evidence for
a structural component is very weak, it is not included in the fit.
This is a reasonable approach, because features with
low surface brightnesses are not expected to have any effect on the 
$B/T$-ratio. In the following the decompositions for some individual galaxies
are discussed.

{\bf NGC718}: Bulge/disk/bar decomposition was made, which gives a
reasonable solution even if the central elliptical structure is not taken 
into account.
In principle this solution gives an upper limit to the $B/T$ light ratio, 
although most probably the central elliptical has only a minor effect on the
relative mass of the bulge. In order to test
the influence of the bar to the $B/T$-ratio, both a Ferrers's and a S\'ersic's 
function were applied
to the bar. The largest relative bulge mass was obtained using a flat Ferrers 
function ($n_{bar}$ = 1), which gave $B/T$ = 0.23, whereas 
Ferrers's function with 
$n_{bar}$ = 4 gave a similar relative bulge mass as obtained 
using a S\'ersic's 
function for the bar ($B/T$ = 0.20 and $B/T$ = 0.18, respectively).
In any case the type of function used to fit the bar appeared to have only a 
minor effect on the obtained $B/T$ light ratio. 
The surface brightness profile of this galaxy looks interesting,
because it shows an exponential-like behavior only in the outer part of 
the galaxy, whereas the surface brightness declines at the radius where 
the bar ends (at $r$ = 25 arcsec). The non-exponential disk is visible 
also in the residual image as negative pixels. The residual image also
shows the flux of the spiral arms above the disk.

{\bf NGC 936}: A decomposition was made which included a bulge, disk,
bar and an oval. The fit to the surface brightness profile is quite 
good, but the residual image still shows some substructure with 
positive and negative pixels in the bar region. 
The substructure at the end of the bar is due to the flux condensation
at the two ends of the bar. The positive residuals at the outer edge
of the disk indicate that the exponential disk fit is failing at
large radii.

{\bf NGC 1400}: The surface brightness profile of this galaxy is relatively 
simple, but no unambiguous solution was found because the fit did not converge.
After several iterations the solution was changed only very slowly, until 
the disk was rapidly lost. However, the solution obtained by this manner 
was unsatisfactory, 
leaving a very large residual in the outer parts of the galaxy.
The decomposition is very similar to the test case shown in Fig 3b, where 
the disk component was also rapidly lost after several iterations.
The best solution was found by stopping the iteration 
slightly before the disk was lost, in which case the residuals are 
considerably smaller. 
For this galaxy, the solution was improved in the central regions
when the shape parameter for the bulge was also used, indicating that
the bulge is slightly flattened. As mentioned in Section 3.2, this
might be the reason for the large $A_2/A_0$-ratio in the central
part of this galaxy.

{\bf NGC 1415}: All the previously identified structural
components are visible in the surface brightness profile of this galaxy: 
the bar, the inner disk, the bulge and the outer disk,
but the image does not have sufficient depth for reliable fitting of 
the outer disk.
As a compromise the bar and the exponential disk were fitted by a single
function, with the consequence that this decomposition can be used
only as an approximation for the $B/T$-ratio, but not for estimating the 
scalelength of the disk (the disk parameters are not shown in Table 7).

{\bf NGC 1440}: In addition to having a bar this galaxy has also a large
lens, which in this particular galaxy somewhat affects the final 
decomposition. 
We found that including the lens to the decomposition considerably 
improves the final model
and changes the relative mass of the bulge from $B/T$ = 0.19 to $B/T$ = 0.14. 
The residual image shows that both the bar and the lens are fairly well 
subtracted.

{\bf NGC 1452}: Reasonable solutions for the surface brightness profile of 
this galaxy can be found both by
applying a bulge/disk/bar or a bulge/disk/bar/oval decomposition.
However, the model image is considerably improved if the oval is also
included. Also, the exponent of $n_{bar}$ = 1 for the bar model gave the 
best fitting solution for the bar. 
The best fit was obtained by finding first a model for the
thin rectangular bar, and then the parameters for the bulge and the oval 
were left free for fitting. This is one of the galaxies that seems to have
a non-exponential disk under the bar region: it appears as a decline
in the surface brightness profile, and also, as positive residuals in the
image where the model is subtracted from the original image.

{\bf NGC 2273}: This galaxy has an inner disk with such a
high surface brightness that it can be identified
in the surface brightness profile of the non-rebinned image (not well 
visible in the rebinned image in Fig. 4).
However, due to the degeneracy between the functions describing the bulge 
and the inner disk we did not include it in the final decomposition. 
For this reason, the obtained $B/T$-ratio can be taken as an upper limit.
After subtracting the galaxy model from the observed image some residuals
are left, demonstrating the presence of the spiral arms and an inner ring
surrounding the bar.

{\bf NGC 2460}: Decompositions for this galaxy have been made previously 
by Carollo et al. (1997), by Baggett, Baggett $\&$ Anderson (1998, hereafter 
BBA98) and by Peng et al. (2002),
which offer interesting points of comparison with our decomposition.
BBA98 used a 1D-method with a S\'ersic's function for the bulge,
and applied their method to profiles from photographic plates that 
had a large enough
field of view for reliable estimation of the radial scalelength of the disk. 
They found that the exponential disk is either very weak 
or absent. On the other hand, Carollo et al. used optical HST-images 
and an $R^{1/4}$ law  
function for the bulge: they comment that this galaxy most probably has a small
bulge. The most sophisticated analysis for this galaxy was made
by Peng et al., who applied a multicomponent 2D decomposition method to 
a high resolution HST-image. 
Including a bar, bulge and a disk in the solution, they obtained $B/T$=0.41,
which is very
similar to $B/T$ = 0.38 found in this study. This corresponds to a typical 
ratio for an Sa-type spiral (Simien $\&$ de Vaucouleurs 1986). We also 
fitted the bar, but it appeared to be unimportant for the obtained $B/T$-ratio.

{\bf NGC 2681}: This galaxy has, in addition to a bulge and a disk, 
also 3 bars and a lens, which in principle all could be taken into 
account in the decomposition. We included only the two larger bars 
and the lens in the decomposition. The secondary bar near to the nucleus (visible in the
non-rebinned image) has a surface brightness
that is too close to the surface brightness of the bulge to be taken
into account in the decomposition, so that the functions for the bulge and
the secondary bar became degenerate.

{\bf NGC 2859}: The best model was found by fitting, besides the bulge 
and the disk, also 2 bars and a lens. Owing to the complex structure
of this galaxy the decomposition was performed in steps so that the
radial scalelength of the disk was found first, then the parameters of the 
primary bar and the lens, and finally the parameters
of the bulge and the secondary bar were left free for fitting. 



{\bf NGC 3626}: This galaxy was found to have a bar and an elliptical
inner structure which might be a secondary bar or an inner disk. Reasonable
decompositions can be made either by fitting both components or only 
the bar component, because the inner elliptical structure does not 
affect the obtained $B/T$-ratio. After subtracting the model from
the original image the residuals are small, manifesting mainly the 
spiral structure of this galaxy.

{\bf NGC 3941}: For this galaxy the parameters for the 
disk and the bar were found first, then they were fixed and the
parameters for the bulge and the lens 
were found simultaneously. A test was also made 
of how much omission of the lens might affect the relative mass of the bulge: 
we found that the decomposition including the oval gives $B/T$ = 0.17, while 
the decomposition without a lens gives $B/T$ = 0.22. 
In such a decomposition model the residuals are extremely small.

{\bf NGC 4245}: This galaxy has two bars and a lens. The best solution 
was found in steps by fitting first the bulge, the disk and the bar. However,
it was more difficult to fit the lens, which easily disappeared.
Finally a solution for the lens was found by fixing its flux shortly before 
the component disappeared. 
The lens in this galaxy has a low surface brightness and does not affect
the obtained $B/T$-ratio. The model-subtracted image shows positive residuals
at the ends of the bar, hinting at the presence of ansae. 

{\bf NGC 4340}: This is again an example of a galaxy with a complicated 
structure having 2 bars and a lens, which were all fitted in the
final decomposition. As usual, the solution was found in steps
starting from the parameters of the disk and the primary bar, and then finding
the parameters for the bulge and the secondary bar. Again, the ansae-type
morphology of the bar is visible in the residual image.

{\bf NGC 4643}: For this galaxy the best decomposition included, besides 
the bulge and the disk, also a bar and an oval. The 
bar model was improved using $n_{bar}$ = 1.
Also, automatic fitting created a slightly too long bar, so that
the bar was forced to have a length that gave the smallest residuals. 
The automatic procedure easily makes a too long bar  
if there is an inner ring surrounding the bar.

It is difficult to compare our results with the previous studies
mainly because in most cases different decomposition methods have been used. 
To our knowledge the only previous studies where multicomponent 
2D decompositions have been used are the studies by 
Prieto et al. (2001) and Peng et al. (2002). With Prieto et al.
we have no galaxies in common, whereas with Peng et al. there is 
one galaxy, NGC 2460, in common. As discussed above, for this galaxy a very 
similar $B/T$-ratio was found in both studies ($B/T$ = 0.38 and 0.41).
A 2D decomposition method has been recently applied for S0s also by 
de Souza, Gadotti $\&$ dos Anjos (2004, hereafter SGA04) using a S\'ersic's 
function for the bulge, but without including a bar in the fitted model. 
We have no galaxies in common with SGA04, but the method is similar as
used by GS03 for NGC 4608. For this galaxy GS03 
found a massive bulge but no clear sign of the disk component, whereas 
we found only a small bulge and a prominent disk
with $B/T$ = 0.15 (see a detailed discussion in Section 5.2).

Another way of evaluating our decompositions is to compare the 
scalelengths of the disks with those obtained in other studies.
We compare $h_r$ estimated by us in the $K_s$-band  
with those obtained by BBA98 in the $V$-band (see Table 8 and Fig. 6). 
BBA98 have measurements for 16 galaxies common with our
sample, of which 13 show reasonable agreement with our measurements.
This is the case in spite of the fact that BBA98 
used a 1D decomposition method, fitted a de Vaucouleurs's $R^{1/4}$ function 
for the bulge, and used a truncated exponential for the disk. However, for NGC 936 
we found a larger $h_r$ than BBA98 (36.8 v.s. 23.8), 
and for NGC 2911, NGC 2859 and NGC 4340 completely
different $h_r$ values were obtained (41.6 vs. 19.2; 78.2 vs. 18.8; 41.1 vs. 
10.9, respectively). These differences seem to be more related
to the depth of the images than to the difference in the decomposition 
method used: most probably for NGC 2859 BBA98
are looking at only the disk in the bar region, whereas our image shows also
the outer disk. For NGC 4340 the reason for the different results 
between the two 
studies is not clear: a comparison with the Digitized Sky Survey image shows 
that for this galaxy our image is not very deep. In any case the very 
small $h_r$ obtained
by BBA98 must be an underestimate. Our measurements are
uncertain for the galaxies that have extremely faint outer disks or 
prominent outer
rings dominating the disk, which is the case for NGC 2781 and NGC 2859.

\section{Discussion of the decompositions}

Two main topics will be discussed in the following: (1) we compare our results
for the $B/T$-ratios and the parameters of the bulge 
with those presented earlier in the literature    
and (2) discuss the possibility raised 
by GS03, that some barred S0s might lack the 
disk component. Particularly the galaxy NGC 4608 will be analyzed.

\subsection{B/T-ratios and the bulge shapes}

Structural decompositions for S0s have repeatedly pointed to a mean $B/T$-ratio
of nearly 0.6 in the $B$-band, as reported also in the recent review by 
Fritze $\&$ Alvensleben (2004).
This is a considerably larger value than $<B/T>$ = 0.41 obtained by 
Simien $\&$ de Vaucouleurs (1986)
for Sa-galaxies, which has been argued to 
indicate that S0s have relative bulge masses intermediate between 
ellipticals and
early-type spirals. The $B/T$-ratio for S0s seems to depend very little 
on the type
of function used for fitting the bulge: for example the de Vaucouleurs's 
$R^{1/4}$ law has been used
in the early studies by Burstein (1979), Simien $\&$ de Vaucouleurs (1986) 
and by Kent (1985), whereas the more general S\'ersic's function has been 
recently used by SGA04, all
giving similar $B/T$-ratios for S0s. The studies by Burstein and 
Simien $\&$ de Vaucouleurs were performed using 
$B$-band images, whereas Kent and SGA04 used $R$-band images. The 
results in the $R$-band can be 
converted to correspond to the $B$-band measurements by applying 
correction terms based on the theoretical models by Schulz et al. 
(2003, hereafter SFAF03). The corrections are based on the 
assumption that the star formation time scales are 
different for bulges and disks, and then spectral evolution in the wavelength
range from UV to $I$-band is modeled in both components 
during the age of the galaxy. Different ages for the bulges and the 
disks are assumed, but the corrections
for S0s are not much different if they have similar or different ages.
In the $R$-band Kent found $<B/T>$ = 0.69, while SGA04 found $<B/T>$ = 0.64. 
After applying the wavelength dependent correction these values are 
converted to correspond to the $B$-band values $<B/T>$ = 0.59 and 0.54, 
respectively, which are very similar to
$<B/T>$ = 0.57, obtained by Simien $\&$ de Vaucouleurs in $B$-band. 
There is only one study, by APB95, 
where a lower $<B/T>$ = 0.32 has been reported in the $K$-band for 
a small number of S0s. Also, for S0s in clusters a somewhat
lower $<B/T>$ = 0.45 has been recently reported by Christlein $\&$ 
Zabludoff (2004, hereafter CZ04) in $R$-band. In the studies where a S\'ersic's 
function is used 
for the bulge, the shape parameter $\beta$ of the bulge is found to be almost identical 
with that corresponding to the $R^{1/4}$ law, which
also explains why the $B/T$-ratios obtained in different studies 
can be so similar.

In this study we found  $<B/T>$ = 0.24 $\pm$ 0.11 for S0s (14 galaxies) 
and $<B/T>$ = 0.28 $\pm$ 0.14 for Sa - S0/a galaxies (9 galaxies). 
The models by SFAF03 do not extend to the near-IR, but the $B/T$-ratios 
at $K_s$-band
are expected to be more similar to those obtained in the $R$-band than 
those obtained in the $B$-band. 
The values found in this study are considerably smaller than reported in 
any of the previous studies, and are likely due to the fact that we have 
used a {\it multicomponent} decomposition method to estimate the $B/T$-ratios. 
The bulges in S0s were also found to be much more exponential-like 
than generally assumed: we found $<n>$ = 2.1 $\pm$ 0.7 (or $\beta$ = 0.48), 
compared 
to $<n>$ = 4.1 (or $\beta$ = 0.24),
given by SGA04 and APB95. The largest $n$-value in our sample is 3.4 
indicating 
that none of the galaxies has a de Vaucouleurs's type surface brightness 
profile. This is in agreement with Balcells et al. (2003), who came to 
the same conclusion using HST-images for their analysis.
The mean $B/T$-ratios for early-type disk galaxies in different studies are
collected in Table 8, where the numbers in parentheses indicate the number 
of galaxies used in the statistics. 

A question then arises why does our study provide such a different results 
compared to 
the previous studies? Our measurements for the $B/T$-ratios and $n$-values 
are compared with those obtained by APB95 and SGA04 in Fig. 7. It is 
particularly 
interesting to compare with the measurements by APB95, because the Kent's 
method for estimating the $B/T$-ratios 
is independent of any assumptions for the model functions. 
It was applied only for non-barred galaxies, for which this method should
work well. It appears that for a given galaxy luminosity the $B/T$-ratios 
found by APB95 are very similar to those obtained in this study, whereas 
our $n$-values are systematically smaller. APB95 had originally a 
magnitude limited sample of 30 S0-Sbc galaxies (of which 7 were S0s), 
of which a subsample of 19 galaxies (including all morphological types) 
was selected for HST imaging (Balcells et al. 2003). For these galaxies 
Balcells et al. then combined the ground-based
images with the resolution of 0.29 arcsec pixel$^{-1}$ with the HST images. 
They applied a decomposition method using a S\'ersic's function for the 
bulge and another function for a possible nucleus and found $<n>$ = 1.7, 
which is considerably smaller than $<n>$ = 3.0 using their ground-based 
images alone. However, a comparison to SGA04 
in Fig. 7 shows that both the $B/T$-ratios 
and the $n$-values found in this study are systematically smaller with 
practically no overlap between the two studies. This is not a luminosity 
effect: the parameters of the bulge are not strongly correlated with galaxy luminosity,
and we also have a large overlap in galaxy luminosity. 
For a given luminosity the $n$-values obtained by SGA04 are also larger
than those by APB95.

The reason for the small $B/T$-ratios and $n$-values found in this 
study are most likely related to the decomposition method, which 
takes into account not only the bulge and the disk, but also bars 
and ovals, which might otherwise be erroneously mixed with the flux 
of the bulge. Indeed, if the fluxes of the bars/ovals were added to the 
flux of the bulge, we would obtain B/T=0.41, which is expected
to mimic the previous bulge/disk decompositions for early-type galaxies.
Other minor factors that might affect the B/T-ratio
are the image resolution and the weighting function applied for noisy images. 
For noisy images mixing
with the bulge might occur particularly if a weighting function giving more 
weight to the inner parts of the galaxy is used. It was pointed out 
by the referee
that probably the largest uncertainty in our multicomponent decomposition
method is the fact that the bar morphologies are not well known, which
might partly explain the low $B/T$-ratios found in this study. Indeed,
boxy/peanut shaped bars are found in galaxies when 
viewed edge-on (L\"utticke, Dettmar $\&$ Pohlen 2000), and some bars
particularly in early-type galaxies have ansae-type morphologies,
where the surface brightness increases towards the outer edges of the bar.
Theoretical models based on N-body simulations  
(Athanassoula 2005 and references therein) also produce this kind of bar:
in these models the inner parts of the bars originally have  
nearly the same vertical thicknesses as the outer parts of the bars,
but the vertical thickness in the inner bar regions increases in time.

In our sample there are many galaxies having bars with ansae-type 
morphologies. In the decompositions they appear as faint 
positive residuals at the two ends of the bar (Fig. 5). 
These structures are faint and their flux is also later added to the 
flux of the disk, so that they are not expected to affect the obtained 
$B/T$-ratio. As the galaxies in our sample are not edge-on systems, it 
might be difficult to distinguish possible thick inner portions of the 
bars from other thick components like ovals/lenses. However, this is not 
critical for our purposes (for $B/T$-ratio), because in both cases
we have used two different functions in the decompositions. However, if
we for some reason have missed the thick part of the bar, that would mean 
that we have underestimated the bar flux and therefore also overestimated 
the $B/T$-ratio. 

There still remains the uncertainty that we have interpreted a bar seen end-on
as a small bulge, which in the simulation models by Athanassoula (2005)
has the properties of a classical bulge. However, as bars are randomly 
oriented in space such cases should be rare. Another
possible uncertainly is the surface brightness profile of the 
primary bar in the presence of the secondary bar: does it have a truncated 
inner structure, induced by secular evolutionary processes in the galaxy?
The present simulation models do not give an answer to this question,
but even if the primary bars were assumed to have a truncated profile,
it is not expected to have any important implications to the B/T flux ratio.
For example, NGC 2859 has both a primary bar and a secondary bar. If the 
secondary bar is included to the decomposition $B/T$ = 0.27, and if it is
not included, $B/T$ = 0.30. As the surface brightness of the
primary bar in this particular galaxy is 2 magnitudes fainter than that of the 
secondary bar, excluding its flux in the region of the secondary bar cannot 
have any effect to the $B/T$-ratio. In our sample we have 4 galaxies
with detectable secondary bars. For two of them (NGC 718, 2681) it was
not taken into account in the decomposition, and two of them (NGC 2858, 
4340) the effect of the secondary bar to the $B/T$ flux ratio was minimal.

\subsection{NGC 4608: a barred S0 without a disk?}

NGC 4608 is one of the two galaxies which have been discussed as 
examples of S0s having prominent bars, but no clear sign of an underlying 
disk (GS03) \footnote{${}^{1}$}{This interpretation has been changed by SGA04:
in the discussion part of their paper the disk of NGC 4608 is considered
to be absent only within the radius of the bar.}.
Structural
decompositions for this galaxy has been made by GS03 both in the optical
and in the near-IR, of which the decomposition in the $V$-band was shown 
by Gadotti $\&$ Souza (2002, their Fig. 1). They used a 2D decomposition 
method 
with an exponential function for the disk and a S\'ersic's function 
for the bulge. The resolution in their $V$-band image 
is 0.3 arcsec pixel$^{-1}$ and the fitting region $r_{max}$ = 135 
arcsec, which can be estimated from their figure. As we also have a $V$-band 
image for this galaxy, having a slightly better resolution of 0.19 
arcsec pixel$^{-1}$, we try to verify their result first.

Bulge/disk decomposition was performed for the rebinned $V$-band image 
using an 
exponential function for the disk and a S\'ersic's function for the 
bulge (Fig. 8a). By limiting the maximum fitting region to 
$r_{max}$ = 135 arcsec, we found a solution having a relatively small 
bulge with $B/T$ = 0.38. Possible effects of the weighting function for 
the solution were investigated. We used the different weighting functions 
described in Section 4.1 and found that in all cases where a 
convergence resulted, the 
$B/T$-ratio depended only slightly on the  
weighting method: the $B/T$-ratio varied between 0.38-0.42 for 
different weighting functions. In the residual image (bottom panel of Fig. 8a) 
the non-subtracted bar is visible as a black feature, and the white 
region inside
the radius of the bar indicates that too much disk is subtracted
in that region (interpreted as a lack of a disk by GS03). 
The only way of fitting the surface brightness 
profile mostly by one single function (as in GS03) was to give strong 
weight for the inner galaxy region using weighting proportional
to $1/r$. However, in that case the solution did not converge and a 
jump to a single function solution occurred rapidly in a similar manner 
as in our test shown in Fig. 3b. As in the test case, also in this case
the residuals after subtracting the bulge were larger than in the solutions 
where the fit converged. Gadotti $\&$ Souza do not specify the weighting 
function they used for NGC 4608, but the solution is not expected to 
depend strongly on the weighting function used.

Finally the decomposition was improved by including, besides the bulge 
and the disk, also the bar and the lens for which 
Ferrers functions were used (Fig. 8b). In that case we obtained 
$B/T$ = 0.19-0.22 for the
rebinned image, depending on the choice of the weighting function. 
As a further test for the reliability of the solution, the models for the
bar and the lens were subtracted from the original image, and then 
a bulge/disk decomposition was made for this residual
image (Fig. 8c): in that case we found $B/T$ = 0.22, which is the same as 
the value obtained by fitting all three components simultaneously.
It seems that omitting the bar in the decomposition causes
the $B/T$-ratio to be overestimated by about 40 $\%$. The 
solution does not depend on the wavelength used: in the $K_s$-band we
we found $B/T$ = 0.19 for the rebinned image, and 
$B/T$ = 0.15 using the original high resolution image. In these 
decompositions $h_R$ was fixed to the previously
found value, for the bar we used $n_{bar}$ = 2 and for the the 
oval $n_{oval}$ = 3.
In this case the shape parameter $c_{bulge}$ was fixed to give elliptical isophotes,
but all the other parameters like the ellipticity of the bulge, were
left free for fitting. Our solution with a small bulge for 
NGC 4608 also corresponds to the visual impression in the original image: 
the galaxy has a small spherical component in the central part of the 
galaxy and a lens with a different ellipticity at a radius which is 
smaller than the radius of the bar. Even if the lens were interpreted 
as part of the bulge, the bright part of the spheroidal component 
would still be small compared to the size of the bar.

We have displayed the surface brightness profile in such a manner that in 
principle all pixels of the 2D image 
can be shown, which makes it possible to better evaluate the presence of the disk.
The profile (Fig. 8 uppermost row) shows a prominent underlying disk 
which is exponential in the outer parts of the galaxy, but
under the bar region at $r < $ 45 arcsec (in the bar minor axis direction) 
the surface brightness of the disk rapidly declines:
e.g. the lower part of the profile declines, although due to the bar 
the total surface brightness at the same radius increases.   
The non-exponential nature of the disk under the bar region is clearly visible 
also in the residual image where the 
model functions are subtracted from the original image (in all three 
decompositions): inside the radius of the bar too much flux is 
subtracted, being manifested as negative residuals in the image. 
The residual image also shows 
faint ring-like structures outside the bar.
In the outermost part of the image ($r > $ 100 arcsec) subtraction of the 
exponential disk is not
very good anymore, mainly because the
surface brightness profile starts to show signs of outer truncation.
If we assume that the inner disk is truncated inside the radius of r $ < $ 20 
arcsec, which
is the crossing point of the bulge and disk models, the $B/T$ = 0.27 
is obtained.
This is an upper limit for the $B/T$-ratio and, in spite of the non-exponential
nature of the inner disk, most probably an 
overestimation of the $B/T$-ratio.
It is not yet clear whether this kind of non-exponential inner disks are 
typical for S0s.
     
Our conclusion is that, contrary to the previous claim by GS03, NGC 4608 has 
only a small bulge and a prominent disk. The disk is exponential
outside the bar and non-exponential in the bar region, where it is about
0.5 mag fainter than expected if the exponential outer disk is extrapolated
to the bar region. In our decomposition the bulge contribution is already
negligible 
at the distance of the end of the bar ($r$ = 35''), where only the disk and 
the bar are contributing to the observed surface brightness. 
The non-exponential nature of the disk in this galaxy is interesting 
and might be a manifestation of secular evolutionary processes 
as predicted by the theoretical models by Athanassoula (2003) and Athanassoula 
$\&$ Misioritis (2003). GS03, based on finding no disk component in this galaxy, 
reach the same conclusion. 
A similar non-exponential disk under the bar region was found also 
in NGC 718, and probably also in NGC 1452.

\section{Summary and conclusions}

A 2D multicomponent decomposition method has been applied to a sample 
of 24 early-type disk galaxies using deep high resolution $K_s$-band 
images obtained at the NOT. The sample,
consisting of 14 S0s and 9 early-type spirals, is part of our magnitude 
limited NIRS0S sample of 170 early-type disk galaxies. The decompositions were
performed using an exponential disk and a S\'ersic's function for the bulge,
while bars and ovals were fitted either using a S\'ersic's or Ferrers's 
functions.
The effect of $seeing$ was taken into account by convolving the model images
with a Gaussian PSF, and the method also allowed deviations from elliptical
isophotes. Finally the decomposition method could be succesfully applied 
only for 22 galaxies: NGC 3414 appeared to be a misclassified edge-on 
galaxy, and for NGC 1415 the decomposition was too uncertain for 
reporting the B/T-ratio in a reliable manner.
In order to avoid non-physical solutions for some S0s with
complex morphological structures, {\it a priori} identification of the 
structural  components was made
by inspecting the radial profiles of the orientation parameters, derived by
fitting ellipses to $B$-band images,  
and by inspecting the radial profiles of the Fourier 
amplitudes and phases. Also, in order to 
distinguish secondary bars from inner disks, unsharp masks were
created, which work as a filter suppressing the low frequency variations 
in the images. Inner spiral arms, not reported previously in the 
literature, were found for the galaxies NGC 1415 and NGC 3941.

The decomposition method was first tested by creating synthetic images 
imitating 
real galaxies, for which the decomposition method was applied in a similar 
manner as for the science images. Particularly we tested how much omission
of a prominent bar affects the obtained $B/T$-ratio, how important is 
the weighting function used, and what are the effects of $seeing$ in the 
decomposition. We found that, by omitting a prominent bar
in the decomposition, the relative mass of the bulge may be overestimated by 
as much as 30-40 $\%$. The exact choice of the weighting function was 
not found to be important unless the images were extremely noisy or if
simple two-component decompositions were applied for 
galaxies with prominent bars. The effects of $seeing$ can be 
properly corrected 
in our method once the PSF in the image is well known. However, if no 
$seeing$ correction is made the $\beta$ parameter of the bulge will be 
overestimated and the profile that originally
had a de Vaucouleurs's type profile starts to approach an 
exponential profile. 

Our most important result is that S0s have bulges that have much smaller
relative masses than previously assumed. We found 
$<B/T>_{K}$ = 0.24 $\pm$ 0.11 (N = 14), compared to $<B/T>_{R}$ = 0.6 
as reported in most previous studies. The most likely reason for the 
low B/T-ratio found in this study is our improved decomposition method, which takes
into account, not only the bulges and disks, but also bars and ovals. Indeed,
if the flux of the bars and ovals were included to the flux of the bulge,
we obtain <B/T>=0.41, which would be close to the value found previously for
early-type galaxies in clusters (CZ04).
It also appeared that the bulges
of S0s do not follow the $R^{1/4}$ law profiles as 
generally assumed. Instead we found $<n>$ = 2.1 $\pm$ 0.7 
(or $\beta$ = 0.48), compared 
to $<n>$ = 4.1 (or $\beta$ = 0.24) found for example by SGA04 and APB95. 
We did not find examples of barred S0s lacking the disk component.
Two such cases had been previously discussed by GS03, of which NGC 4608 was  
analyzed in this study. For this galaxy we found
$B/T$ = 0.38, if only the bulge and the disk were taken into 
account in the decomposition, and more realistically $B/T$ = 0.15 if the bar
and the lens were also included in the fit. This galaxy was also found to have
a non-exponential inner disk, which might be a manifestation
of secular evolution as predicted by the models by Athanassoula (2003) and 
Athanassoula $\&$ Misioritis (2003). Another similar case was NGC 718, 
and possibly also NGC 1452.
Quite interestingly S0/a - Sa galaxies in our sample were found to have
similar $B/T$-ratios ($<B/T>_{K}$ = 0.28 $\pm$ 0.14 (N = 9)) as the S0s.
To our knowledge our study is the first attempt to apply 2D multicomponent 
decomposition method for a medium sized sample of early-type disk galaxies.


\section*{Acknowledgments}

We thank the anonymous referee for a careful reading of the manuscript.
EL and HS acknowledge the support of the Academy of Finland, and EL also 
from the Magnus Ehrnrooth Foundation.
This research has also made use of the NASA/IPAC Extragalactic
Database (NED) which is operated by the Jet Propulsion 
Laboratory, California Institute of Technology, under
contract with the National Aeronautics and Space Administration.
RB also acknowledges the support of NSF grant AST 0205143 to the 
University of Alabama.

\vfill
\eject

\subsection{Figure captions}

{\bf Figure 1:} Polar angle maps of all $K_s$-band images of our sample.
The x-axis is the angle from the line of nodes counted along the disk plane,
and the y-axis is the distance from the galaxy center in arcseconds.
In order to better illustrate the non-axisymmetric features, the axisymmetric
m = 0 components are subtracted from the original images. We also show the
m = 2 and m = 4 Fourier amplitudes of density, normalized by the m = 0 component,
and their phases. The images we use were deprojected to face-on orientation
as explained in Section 3.2.
\vskip 0.25cm

{\bf Figure 2:} {\it Test a} shows bulge/disk decomposition for a
synthetic image having an exponential disk with $h_r$ = 45 and a 
S\'ersic bulge with $\beta$ = 0.6 and $h_b$ = 1. In order 
to imitate realistic galaxy images, Poisson noise and background noise were 
also added.
{\it Tests b} and $c$ show 
decompositions for a synthetic image in which bar component, described
by a Ferrers function, was also added. In {\it test b} only the 
bulge and the disk were fitted, whereas in {\it test c} 
all three components were fitted simultaneously.  
In the surface brightness profile, each pixel in the image is shown 
as a function of its distance
from the center measured in the sky plane ({\it black dots}). Also shown are 
the model components for the bulge and the disk ({\it light grey}),
for the bar ({\it darker grey}), and for the total model ({\it dark grey}). 
For each decomposition, the synthetic image, model 
image, and residual (difference) image are also shown.
\vskip 0.25cm

{\bf Figure 3:} Decompositions were made for three synthetic images with 
a Sersic's bulge and an exponential disk with $h_R$ = 50. As in Fig. 2
Poisson noise and background noise were 
also added. The two synthetic images in tests $a$ and $b$ were taken to have 
nearly $R^{1/4}$ like total surface brightness profiles, but different 
properties of the bulge, as indicated in the figure. The third synthetic 
image was used in {\it tests c} and {\it d}, having a small bulge and
a large and noisy disk ($h_R$ = 50). Two decompositions were made for this image,
which were otherwise similar but deviated in the choice of the weighting 
function:
in {\it test d} pixels were weighted in proportion to $1/F_i$, whereas in
{\it test c} they were weighted in proportion to $r_i$.
\vskip 0.25cm

{\bf Figure 4:} Multicomponent 2D decomposition applied to 
$K_s$-band images for all galaxies in our sample. 
In order to better illustrate the decompositions, the images were first 
rebinned by a factor of 4. Notice however, that in Table 6, where the 
final results are given, non-rebinned images were used. For each galaxy 
we show the observed image ({\it upper left panel}), the image
constructed from the model functions, shown in the
same orientation ({\it lower left panel}), and the surface brightness profile 
together with the fitted model functions ({\it right panel}). The colors 
in the profiles are the same as in Fig. 2, except that more components 
are shown.  The zero point of the surface brightness is arbitrary, but one
unit corresponds to one magnitude.
\vskip 0.25cm

{\bf Figure 5:} Residual images for all galaxies. In the {\it left 
panels} are shown the residual images from the decompositions, obtained 
by subtracting the model images from the observed images. In the 
{\it middle panel} are shown the unsharp masks, created by subtracting 
a heavily smoothed copy from the original image. The unsharp masks
(positive parts) are shown in a logarithmic scale with contours over-plotted 
on the images for better illustrating the faint structural components 
in the galaxies. The x- and y-coordinates in both images are given in 
arcseconds. In the {\it right panel} are also shown the radial profiles
of the position angles $\phi$ and the minor-to-major axis ratios $q$ for 
these galaxies, shown on a logarithmic radial scale. The residual images 
are obtained using the $K_s$-band images, whereas the radial profiles 
for the orientation parameters were derived mainly using the $B$-band 
images (see Table 2). Notice that for the unsharp mask images, only the 
central parts of the images are shown.
\vskip 0.25cm

{\bf Figure 6:} The scalelengths of the disks obtained in this study 
in the $K_s$-band are compared with those obtained by Baggett, Baggett $\&$ 
Anderson (1998) in the $V$-band (NGC 2859 is outside of the figure). 
The line is a unit slope. 
\vskip 0.25cm

{\bf Figure 7:} The parameters of the bulge found in this study 
are compared with those obtained by Andredakis,
Peletier $\&$ Ballcels (1995, APB) who used a modified 
Kent's method applied in the $K_s$-band, and with those obtained by de 
Souza, Gadotti $\&$ dos Anjos (2004, SGA) in the $R$-band  
using a 2D method with an exponential function for the disk and a 
S\'ersic's function for the bulge. 
The {\it thick crosses} show our measurements for S0s and 
the {\it thin crosses} are our measurements for S0/a - Sa galaxies. 
The {\it triangles} indicate the measurements by APB, 
whereas the {\it squares} show the measurements by SGA. 
The {\it upper panel} shows the shape parameter of the bulge, $n$, as 
a function
of the $B/T$-ratio, and in the two lower panels $B/T$ and $n$ are shown
as a function of galaxy luminosity. The estimated uncertainty in the $n$
parameter in our study is 20\%, based on the evaluation of the effect
of the weigting function in the decomposition applied for noisy test images.
The $B$-band magnitudes used to calculate the absolute magnitudes are from RC3.
Note that the dispersion is large in the two lower panels, where 
the dependence of these parameters of the galaxy magnitude was studied.
\vskip 0.25cm

{\bf Figure 8:} Structural decompositions for NGC 4608 in the $V$-band. 
The left column ({\it Fig.  8a}) shows the bulge/disk decomposition,
whereas in the middle column ({\it Fig. 8b}) the bar component 
is also added to the fit. 
{\it Fig. 8c} shows a decomposition made for the residual image obtained 
after subtracting 
the fitted bar/oval models from the original image. 
For all decompositions, the original, model, and residual images are also 
shown.

\vfill
\eject

\begintable*{1}
\caption{{\bf Table 1.} Observations.}

\halign{%
\rm#\hfil&\qquad\rm#\hfil&\qquad\rm\hfil#&\qquad\rm\hfil
#&\qquad\rm\hfil#&\qquad\rm\hfil#&\qquad\rm#\hfil
&\qquad\rm\hfil#&\qquad\rm#\hfil&\qquad\hfil\rm#\cr
\noalign{\vskip 3pt\hrule\vskip 3pt}
          &              &         &         &             \cr
Galaxy    & RC3 type      &filter   & FWHM    & nuclear act.  \cr
          &              &         & [arcsec]&               \cr
\noalign{\vskip 5pt}
\noalign{\vskip 3pt\hrule\vskip 3pt}
\noalign{\vskip 5pt}

NGC 718  & SAB(s)a      & $K_s$   & 1.06   &       \cr
         &               & $B$     & 1.29   &      \cr
NGC 936  & SB(rs)0$^+$       & $K_s$   & 1.24   &       \cr  
         &               & $B$     & 1.31   &       \cr
NGC 1022 & (R$^{\prime}$)SB(s)a       & $K_s$   & 1.84   & HII/SB      \cr
         &               & $B$     & 0.99   &      \cr
NGC 1400 & SA0$^-$        & $K_s$   & 1.06   &      \cr
         &               & $B$     & 1.25   &     \cr
NGC 1415 & (R)SAB(s)0/a       & $K_s$   & 1.15   &     \cr
         &               & $B$     & 1.20   &     \cr
NGC 1440 & (R$^{\prime}$)SB(rs)0$^{\circ}$:       & $K_s$   & 0.99   &     \cr
NGC 1452 & (R$^{\prime}$)SB(r)0/a       & $K_s$   & 1.04   &     \cr
         &               & $B$     & 1.41   &     \cr
NGC 2196 & (R$^{\prime}$)SA(s)a       & $K_s$   & 1.40   &     \cr
         &               & $B$     & 1.16   &     \cr
NGC 2273 & SB(r)a:        & $K_s$   & 1.31   &     \cr
NGC 2460 & SA(s)a       & $K_s$   & 0.99   &     \cr
NGC 2681 & (R$^{\prime}$)SAB(rs)0/a       & $K_s$   & 1.15   & Sy     \cr
         &               & $B$     & 1.10   &      \cr
NGC 2781 & SAB(r)0$^+$       & $K_s$   & 0.90   &      \cr
NGC 2855 & (R)SA(rs)0/a       & $K_s$   & 0.92   &     \cr 
         &               & $B$     & 1.39   &      \cr
NGC 2859 & (R)SB(r)0$^+$       & $K_s$   & 0.92   & Sy      \cr
         &               & $B$     & 0.82   &      \cr
NGC 2911 & SA(s)0: pec       & $K_s$   & 0.78   & Sy/L     \cr
NGC 2983 & SB(rs)0$^+$       & $K_s$   & 0.97   &      \cr
NGC 3414 & S0 pec      & $K_s$   & 1.01   &      \cr
         &               & $B$     & 1.14   &      \cr
NGC 3626 & (R)SA(rs)0$^+$       & $K_s$   & 0.83   &      \cr
NGC 3941 & SB(s)0$^{\circ}$        & $K_s$   & 0.87   & Sy2     \cr
NGC 4245 & SB(r)0/a:       & $K_s$   & 0.99   & HII     \cr
         &               & $B$     & 0.91   &      \cr
NGC 4340 & SB(r)0$^+$       & $K_s$   & 1.04   &      \cr
         &               & $B$     & 1.88   &      \cr
NGC 4596 & SB(r)0$^+$       & $K_s$   & 0.85   & L      \cr
         &               & $B$     & 1.31   &      \cr
NGC 4608 & SB(r)0$^{\circ}$       & $K_s$   & 1.10   &     \cr
         &               & $B$     & 0.89   &     \cr
NGC 4643 & SB(rs)0/a       & $K_s$   & 0.97   & L     \cr
\noalign{\vskip 5pt}
\noalign{\vskip 3pt\hrule\vskip 3pt}
}
\endtable

\vfill
\eject

\begintable*{2}
\caption{{\bf Table 2.} The orientation parameters }
\advance\baselineskip by 2pt
\halign{#\hfil\quad&#\hfil\quad&#\hfil\qquad\qquad
        &#\hfil\quad&#\hfil\quad&#\hfil
        &#\hfil\quad&#\hfil\quad&#\hfil
        &#\hfil\quad&#\hfil\quad&#\hfil\cr
\noalign{\vskip 3pt\hrule\vskip 3pt}
          &                  &                       &          &          &              &            &              &        \cr
Galaxy    & $\phi$(measured) &  $q$ (measured)       & range    & filter   &$\phi$ (RC3)  &$q$ (RC3)   &$\phi$ (ES03) &$q$ (ES03) \cr
          &  [$^o$]          &                       & [arcsec] &          &   [$^o$]     &            & [$^o$]       &        \cr
 (1)      & (2)              &  (3)                  &  (4)     &  (5)     &  (6)         &   (7)      &   (8)        & (9)  \cr   
\noalign{\vskip 5pt}
\noalign{\vskip 3pt\hrule\vskip 3pt}
\noalign{\vskip 5pt}

          &                  &                       &          &          &              &            &            &         \cr
NGC 718  &   7.4  $\pm$ 2.5  &  0.852  $\pm$ 0.014   &  80-95   & $B$      &    45        & 0.871 & 5    & 0.87 \cr
NGC 936  & 122.7  $\pm$ 1.6  &  0.739  $\pm$ 0.009   &  170,180 & $B$      &              & 0.871 & 130  & 0.76 \cr
NGC 1022 &  26.7  $\pm$ 2.5  &  0.930  $\pm$ 0.016   &  88-115  & $B$      &              & 0.832 & 174  & 0.92 \cr
NGC 1400 &  37.1  $\pm$ 5.3  &  0.915  $\pm$ 0.013   &  100-120 & $B$      &    40        & 0.871 &      &      \cr   
NGC 1415 & 151.7  $\pm$ 0.5  &  0.390  $\pm$ 0.005   &  160-180 & $B$      &   148        & 0.513 &      &     \cr 
NGC 1440 &  22.2  $\pm$ 2.5  &  0.777  $\pm$ 0.026   &   60-70  & $K_s$    &              & 0.759 &      &      \cr
NGC 1452 & 113.4  $\pm$ 0.8  &  0.698  $\pm$ 0.010   &   90-100 & $B$      &   113        & 0.661 &      &      \cr
NGC 2196 &  67.8  $\pm$ 1.5  &  0.697  $\pm$ 0.012   &  130-140 & $B$      &    35        & 0.776 &      &      \cr
NGC 2273 &  59.3  $\pm$ 1.2  &  0.599  $\pm$ 0.016   &  65-68   & $K_s$    &              & 0.759 &  50  & 0.66  \cr
NGC 2460 &  29.5  $\pm$ 3.1  &  0.750  $\pm$ 0.027   &  45-55   & $K_s$    &              & 0.759 &      &       \cr
NGC 2681 & 102.0  $\pm$ 5.9  &  0.912  $\pm$ 0.017   &   120-145& $B$      &              & 0.912 &  140 & 0.95  \cr
NGC 2781 &  72.0  $\pm$ 2.0  &  0.515  $\pm$ 0.012   &   45-55  & $K_s$    &              & 0.501 &      &       \cr
NGC 2855 & 107.2  $\pm$ 2.7  &  0.838  $\pm$ 0.011   &   140-150& $B$      &   130        & 0.891 &      &       \cr
NGC 2859 &  86.3  $\pm$ 1.3  &  0.760  $\pm$ 0.012   &   160-175& $B$      &    85        & 0.891 & 90   & 0.90  \cr
NGC 2911 & 133.4  $\pm$ 3.6  &  0.763  $\pm$ 0.021   &   30-35  & $K_s$    &              & 0.776 &      &       \cr
NGC 2983 &  91.1  $\pm$ 0.9  &  0.578  $\pm$ 0.016   &   65-70  & $K_s$    &              & 0.589 &      &       \cr
NGC 3414 &  22.7  $\pm$ 2.7  &  0.856  $\pm$ 0.014   &   110-120& $B$      &              & 0.724 &      &       \cr
NGC 3626 & 158.9  $\pm$ 0.8  &  0.663  $\pm$ 0.011   &   70-73  & $K_s$    &              & 0.724 &      &       \cr
NGC 3941 &   6.9  $\pm$ 0.8  &  0.651  $\pm$ 0.013   &   70-80  & $K_s$    &              & 0.661 & 10   & 0.65  \cr
NGC 4245 & 174.1  $\pm$ 2.2  &  0.823  $\pm$ 0.011   &  110-127 & $B$      &              & 0.759 & 0    & 0.75  \cr
NGC 4340 &  98.6  $\pm$ 1.3  &  0.556  $\pm$ 0.012   &  150-165 & $B$      &              & 0.794 &      &       \cr
NGC 4596 & 116.0  $\pm$ 0.5  &  0.716  $\pm$ 0.007   &  160-175 & $B$      &              & 0.741 &      &       \cr
NGC 4608 & 101.0  $\pm$ 3.5  &  0.856  $\pm$ 0.015   &  130-150 & $B$      &              & 0.832 &      &       \cr
NGC 4643 &  49.8  $\pm$ 5.6  &  0.836  $\pm$ 0.023   &  100-105 & $K_s$    &              & 0.741 & 55   & 0.80  \cr
\noalign{\vskip 5pt}
\noalign{\vskip 3pt\hrule\vskip 3pt}
}

\tabletext{\noindent The columns are: (1) the galaxy name, (2, 3) 
the measured position angle $\phi$ and minor-to-major axis ratio $q$, 
(4, 5) the range and filter used for the measurements, (6, 7) $\phi$ and $q$ as given 
in RC3, (8, 9)  $\phi$ and $q$ as given by ES03.} 

\endtable

\vfill
\eject


\begintable*{3}
\caption{{\bf Table 3.} The identified structural components.}


\advance\baselineskip by 2pt
\halign{#\hfil\quad&#\hfil\quad&#\hfil\qquad\qquad
        &#\hfil\quad&#\hfil\quad&#\hfil\cr
\noalign{\vskip 3pt\hrule\vskip 3pt}
          &               &                    \cr
Galaxy    & primary bar   & other components   \cr
\noalign{\vskip 5pt}
\noalign{\vskip 3pt\hrule\vskip 3pt}
\noalign{\vskip 5pt}

NGC 718   & $bar$ (20 arcsec)       &  $bar_2$ or nuclear spirals       \cr
NGC 936   & $bar$ (55 arcsec)       &  nuclear ring (8-9 arcsec), lens   \cr
NGC 1022  & $bar$ (25-30 arcsec)    &  oval, inner pseudoring           \cr   
NGC 1400  &                         &                                   \cr 
NGC 1415  & $bar$ (70 arcsec)       & inner spirals (9 arcsec)          \cr 
NGC 1440  & $bar$ (30 arcsec)       & lens                               \cr
NGC 1452  & $bar$ (46 arcsec)       & oval, ring                         \cr
NGC 2196  &                         & inner elliptical (16 arcsec)      \cr
NGC 2273  & $bar$ (20 arcsec)       & inner spirals (5 arcsec)           \cr
NGC 2460  &                         & $bar_2$ (8 arcsec)                  \cr
NGC 2681  & $bar_1$ (20 arcsec)     & $bar_2$ (5 arcsec), $bar_3$ (60 arcsec), lens  \cr
NGC 2781  & $bar$ (45 arcsec)       & inner spiral arms (10 arcsec)       \cr
NGC 2855  &                         &                                      \cr
NGC 2859  & $bar_1$ (57 arcsec)     & $bar_2$ (8 arcsec), two lenses, outer ring  \cr
NGC 2911  &                         &                                        \cr
NGC 2983  & $bar$ (36 arcsec)       & inner elliptical (5 arcsec), lens        \cr
NGC 3414  &                         & lens                                  \cr
NGC 3626  & $bar$ (40 arcsec)       & inner bar or a disk (5 arcsec), two lenses  \cr
NGC 3941  & $bar$ (30 arcsec)       & inner spirals (4 arcsec),  lens      \cr
NGC 4245  & $bar_1$ (46 arcsec)     & nuclear ring,  inner ring, lens     \cr
NGC 4340  & $bar_1$ (75 arcsec)     & $bar_2$ (10 arcsec), lens           \cr
NGC 4596  & $bar$ (74 arcsec)       & inner elliptical (7 arcsec), lens    \cr
NGC 4608  & $bar$ (55 arcsec)       & inner ring, lens                        \cr
NGC 4643  & $bar$ (60 arcsec)       & possible nuclear ring (10 arcsec), lens  \cr
\noalign{\vskip 5pt}
\noalign{\vskip 3pt\hrule\vskip 3pt}
}
\endtable

\vskip 5cm


\begintable*{4a}
\caption{{\bf Table 4a.} The effect of the weighting method when $S/N=3.0$.}
\halign{%
\rm#\hfil&\qquad\rm#\hfil&\qquad\rm\hfil#&\qquad\rm\hfil
#&\qquad\rm\hfil#&\qquad\rm\hfil#&\qquad\rm#\hfil
&\qquad\rm\hfil#&\qquad\rm#\hfil&\qquad\hfil\rm#\cr
\noalign{\vskip 3pt\hrule\vskip 3pt}
                        &          &             &       &         &  \cr
weight                  & $h_R$    & rel. error  & $B/D$ & $\beta$ & $r_{eff}$   \cr
\noalign{\vskip 5pt}
\noalign{\vskip 3pt\hrule\vskip 3pt}
\noalign{\vskip 5pt}
$1/r^2$                 &  27.96   & -6.80  $\%$ & 0.206 & 0.533  & 2.950     \cr
$1/(F_{model} r^2)$     &  29.83   & -0.56  $\%$ & 0.203 & 0.508  & 2.963      \cr
$1/r$                   &  29.55   & -1.49  $\%$ & 0.202 & 0.515  & 3.005     \cr
$const$                 &  29.91   & -0.30  $\%$ & 0.204 & 0.505  & 3.021     \cr
$1/(F_{model} r)$       &  29.98   & -0.07  $\%$ & 0.205 & 0.502  &  2.999     \cr
$r$                     &  30.00   & -0.01  $\%$ & 0.205 & 0.499  &  3.002     \cr
$1/F_{model}$           &  30.02   & +0.07  $\%$ & 0.205 & 0.495  & 3.004     \cr
$r/(F_{model})$         &  30.05   & +0.15  $\%$ & 0.207 & 0.478  & 2.990     \cr
                        &          &             &       &        &        \cr
$1/(F r^2)$             &  29.35   &-2.18  $\%$  & 0.207 & 0.512  &  3.004    \cr
$1/(F r)$               &  29.55   &-1.50  $\%$  & 0.208 & 0.503  &  2.990     \cr
$1/F$                   &  29.67   &-1.10  $\%$  & 0.210 & 0.493  & 3.009      \cr
$r/F$                   &  29.87   &-0.45  $\%$  & 0.215 & 0.448  & 3.027      \cr
\noalign{\vskip 5pt}
\noalign{\vskip 3pt\hrule\vskip 3pt}
}
\endtable

\vskip 5cm

\begintable*{4b}
\caption{{\bf Table 4b.} The effect of the weighting method:$S/N$=1.5}
\halign{%
\rm#\hfil&\qquad\rm#\hfil&\qquad\rm\hfil#&\qquad\rm\hfil
#&\qquad\rm\hfil#&\qquad\rm\hfil#&\qquad\rm#\hfil
&\qquad\rm\hfil#&\qquad\rm#\hfil&\qquad\hfil\rm#\cr
\noalign{\vskip 3pt\hrule\vskip 3pt}
                        &         &             &       &         &      \cr
weight                  & $h_R$   & rel. error  & $B/D$ & $\beta$ & $r_{eff}$   \cr
\noalign{\vskip 5pt}
\noalign{\vskip 3pt\hrule\vskip 3pt}
\noalign{\vskip 5pt}
$1/r^2$                 &  26.09  & -13.03 $\%$  &0.209 &0.564 &2.921  \cr
$1/(F_{model} r^2)$     &  29.78  &  -0.73 $\%$  &0.203 &0.511 &3.062   \cr
$1/r$                   &  29.03  &  -3.24 $\%$  &0.201 &0.531 &3.007   \cr
$const$                 &  29.82  &  -0.60 $\%$  &0.203 &0.510 &3.045   \cr
$1/(F_{model} r)$       &  29.95  &  -0.16 $\%$  &0.204 &0.503 &3.000   \cr
$r$                     &  30.00  &  -0.00 $\%$  &0.205 &0.496 &3.006    \cr
$1/F_{model}$           &  30.04  &  +0.12 $\%$  &0.206 &0.492 &3.006    \cr
$r/(F_{model})$         &  $-$    &              &      &      &         \cr
                        &         &              &      &      &         \cr
$1/(F r^2)$             &  27.91  &   -7.0 $\%$  &0.216 &0.527 &2.994    \cr
$1/(F r)$               &  28.41  &   -5.3 $\%$  &0.220 &0.506 &2.916    \cr
$1/F$                   &  28.80  &   -4.0 $\%$  &0.225 &0.473 &3.049    \cr
$r/F$                   &  29.41  &   -2.0 $\%$  &0.244 &0.379 &3.237    \cr
\noalign{\vskip 5pt}
\noalign{\vskip 3pt\hrule\vskip 3pt}
}
\endtable

\vskip 5cm

\begintable*{5}
\caption{{\bf Table 5.} The effect of seeing in the estimated $\beta$. Subscript $o$ denotes the actual value in the synthetic image.}
\halign{%
\rm#\hfil&\qquad\rm#\hfil&\qquad\rm\hfil#&\qquad\rm\hfil
#&\qquad\rm\hfil#&\qquad\rm\hfil#&\qquad\rm#\hfil
&\qquad\rm\hfil#&\qquad\rm#\hfil&\qquad\hfil\rm#\cr
\noalign{\vskip 3pt\hrule\vskip 3pt}
                             &                  &                &                  \cr
($\sigma_{PSF}/r_{eff}$)$_o$      & $\beta_0=0.25$   & $\beta_0=0.50$ & $\beta_0=1.00$  \cr
\noalign{\vskip 3pt\hrule\vskip 3pt}
seeing ignored in the fit  &$\beta$ &$\beta$ & $\beta$   \cr
\noalign{\vskip 3pt\hrule\vskip 3pt}
      &         &              &   \cr
0.05  & 0.2825  & 0.5136       & 1.005  \cr
0.10  & 0.3371  & 0.5448       & 1.019  \cr
0.25  & 0.4832  & 0.6638       & 1.102  \cr
0.50  & 0.7095  & 0.8842       & 1.299  \cr
0.75  & 0.9382  & 1.112        & 1.501  \cr
1.00  & 1.160   & 1.330        & 1.669  \cr
      &         &              &         \cr
\noalign{\vskip 3pt\hrule\vskip 3pt}
seeing underestimated by 10\% &$\beta$ &$\beta$ & $\beta$ \cr
\noalign{\vskip 3pt\hrule\vskip 3pt}
      &         &              &     \cr
0.05  & 0.2539  & 0.5020   & 1.001 \cr
0.10  & 0.2594  & 0.5060   & 1.003 \cr
0.25  & 0.2769  & 0.5220   & 1.017 \cr
0.50  & 0.3108  & 0.5577   & 1.055 \cr
0.75  & 0.3528  & 0.6037   & 1.108 \cr
1.00  & 0.4063  & 0.6625   & 1.177 \cr
      &         &          &        \cr
\noalign{\vskip 3pt\hrule\vskip 3pt}
seeing overestimated by 10\% &$\beta$ &$\beta$ & $\beta$ \cr
\noalign{\vskip 3pt\hrule\vskip 3pt}
& & &   \cr
0.05  & 0.2463  & 0.4980   & 0.9990  \cr
0.10  & 0.2410  & 0.4940   & 0.9963  \cr
0.25  & 0.2237  & 0.4777   & 0.9815  \cr
0.50  & 0.1981  & 0.4408   & 0.9397  \cr
0.75  & 0.1914  & 0.3907   & 0.8747  \cr
1.00  & 0.1914  & 0.3318   & 0.7766  \cr
\noalign{\vskip 5pt}
\noalign{\vskip 3pt\hrule\vskip 3pt}
}
\endtable

\vfill
\eject

\begintable*{6}
\caption{{\bf Table 6.} Bulge/disk and bulge/disk/bar decompositions for 3 synthetic images.
The values in parentheses are deviations from the correct values.}
\halign{%
\rm#\hfil&\qquad\rm#\hfil&\qquad\rm\hfil#&\qquad\rm\hfil#
&\qquad\rm\hfil#&\qquad\rm\hfil#&\qquad\rm#\hfil
&\qquad\rm\hfil#&\qquad\rm\hfil#&\qquad\rm#\hfil
&\qquad\rm\hfil#&\qquad\rm\hfil#&\qquad\rm#\hfil
&\qquad\rm\hfil#&\qquad\rm#\hfil&\qquad\hfil\rm#\cr
\noalign{\vskip 3pt\hrule\vskip 3pt}
         &          &               &              &                &               \cr
         &          & bulge/disk    &              & bulge/disk/bar &               \cr
         & weight   & $B/D$ (dev)   & $\beta$ (dev)& $B/D$ (dev)    & $\beta$ (dev)  \cr
        &          &               &              &                &                \cr 
\noalign{\vskip 3pt\hrule\vskip 3pt}
model 1: $B/D$=1.0, $\beta$=0.5, $q$=1.0 & & & & &  \cr
\noalign{\vskip 3pt\hrule\vskip 3pt}
        &          &               &              &                &                \cr 
         & 1/F      & 0.97 (3 $\%$) & 0.50 (0 $\%$)& 1.23 (23 $\%$) & 0.59 (18 $\%$)) \cr
         & r        & 0.98 (2 $\%$) & 0.50 (0 $\%$)& 1.66 (66 $\%$) & 0.52 (4 $\%$)   \cr
         & constant & 0.97 (3 $\%$) & 0.50 (0 $\%$)& 2.37           & 0.46 (8 $\%$)   \cr
         &          &               &              &                &                \cr   
\noalign{\vskip 3pt\hrule\vskip 3pt}
model 2: $B/D$=0.92, $\beta$=0.5, $q$=0.9 & & & & &  \cr
\noalign{\vskip 3pt\hrule\vskip 3pt}
         & 1/ F     & 0.87 (5 $\%$) & 0.50 (0 $\%$)& 1.13 (22 $\%$) & 0.60(20 $\%$)   \cr
         & r        & 0.88 (4 $\%$) & 0.50 (0 $\%$)& 1.50 (63 $\%$) & 0.52(4 $\%$)    \cr
         & constant & 0.88 (4 $\%$) & 0.50 (0 $\%$)&  -             &  -             \cr
         &          &               &              &                &                \cr
\noalign{\vskip 3pt\hrule\vskip 3pt}
model 3: $B/D$=1.0, $\beta$=0.33, $q$=1.0 & & & & &  \cr
\noalign{\vskip 3pt\hrule\vskip 3pt}
        &          &               &              &                &                \cr 
         & 1/F      & 1.00 (0 $\%$) & 0.33 (0 $\%$)& 0.81 (19 $\%$) & 0.53 (60 $\%$) \cr
         & r        & 1.00 (0 $\%$) & 0.33 (0 $\%$)& 0.98 (2 $\%$)  & 0.50 (51 $\%$) \cr
         & constant & 1.00 (0 $\%$) & 0.33 (0 $\%$)& 20.80          & 0.28 (15 $\%$) \cr
\noalign{\vskip 5pt}
\noalign{\vskip 3pt\hrule\vskip 3pt}
}
\endtable

\vfill
\eject

\vskip 5cm

\begintable*{7}
\caption{{\bf Table 7.} The decomposition results. The parameter B(all)/T in the last column indicates the B/T-ratio, when the fluxes
of bars and ovals are included to the flux of the bulge.}
\halign{%
\rm#\hfil&\qquad\rm#\hfil&\qquad\rm\hfil#&\qquad\rm\hfil#
&\qquad\rm\hfil#&\qquad\rm\hfil#&\qquad\rm#\hfil
&\qquad\rm\hfil#&\qquad\rm\hfil#&\qquad\rm#\hfil
&\qquad\rm\hfil#&\qquad\rm\hfil#&\qquad\rm#\hfil
&\qquad\rm\hfil#&\qquad\rm#\hfil&\qquad\hfil\rm#\cr
\noalign{\vskip 3pt\hrule\vskip 3pt}
        &          &               &       &            &           &          &        &           &          &            &     &      &             \cr
        & Bulge    &               &       &Disk        &(BBA)      &Ferr 1    &        &Ferr 2     &          &Ferr 3      &     &$B/T$ &$B(all)/T$ \cr
 NGC    &  $\beta$ & $r_{eff}$[``] & q     &$h_r$ [``]  &$h_r$ [``] & $a$ [``] &  q     &$a$ [``]   & q        & $a$  [``]  & q   &      &                        \cr
\noalign{\vskip 5pt}
\noalign{\vskip 3pt\hrule\vskip 3pt}
\noalign{\vskip 5pt}
  718   & 0.69     & 2.25     & 0.8   &  21.9     & 24.0     &  33.5   &  0.5     &          &          &           &     &  0.21 &   0.34     \cr
  936   & 0.68     & 3.14     & 0.9   &  36.9     & 23.8     &  74.6   &  0.3     &  26.2    & 0.9      &           &     &  0.12 &   0.33      \cr
 1022   & 0.46     & 1.41     & 0.9   &  19.8     & 19.5     &  32.6   &  0.4     &  12.7    & 0.6      &           &     &  0.10 &   0.29      \cr
 1400   & 0.39     & 6.38     & 0.9   &  18.7     &          &         &          &          &          &           &     &  0.52 &   0.52      \cr
 1415   & 0.63     & 1.67     & 0.6   &           & 21.5     &         &          &          &          &           &     &       &              \cr
 1440   & 0.71     & 2.33     & 0.8   &  20.4     & 19.9     &  35.8   &  0.3     &  16.0    & 0.9      &           &     &  0.14 &   0.33     \cr
 1452   & 0.81     & 2.88     & 0.9   &  32.8     & 24.7     &  60.0   &  0.2     &  24.7    & 0.9      &           &     &  0.14 &   0.36      \cr
 2196   & 0.36     & 9.28     & 0.8   &  24.1     & 27.2     &         &          &          &          &           &     &  0.45 &   0.44     \cr
 2273   & 0.55     & 2.64     & 0.6   &  25.0     &          &  52.3   &  0.2     &          &          &           &     &  0.23 &   0.41     \cr
 2460   & 0.39     & 4.69     & 0.8   &  10.5     &          &         &          &  20.5    & 0.4      &           &     &  0.27 &   0.29     \cr
 2681   & 0.46     & 1.82     & 0.9   &  33.3     & 31.0     &  30.0   &  0.7     &  64.4    & 0.7      & 12.6      & 0.8 &  0.24 &   0.45      \cr
 2781   & 0.35     & 4.36     & 0.8   &  54.7     &          &  52.2   &  0.9     &  14.2    & 1.0      &           &     &  0.29 &   0.65      \cr
 2855   & 0.35     & 12.95     & 0.8   &  32.2     &          &         &          &          &          &           &     &  0.57 &   0.57      \cr
 2859   & 0.77     & 4.65     & 0.9   &  78.2     & 18.8     &  79.6   &  0.4     &  10.2    & 0.6      & 73.4      & 1.0 &  0.27 &   0.54      \cr
 2911   & 0.32     & 6.99     & 0.8   &  41.6     & 19.2     &         &          &          &          &           &     &  0.34 &   0.33      \cr
 2983   & 0.76     & 1.70     & 0.8   &  27.6     &          &  47.9   &  0.3     &  17.9    & 1.0      &           &     &  0.11 &   0.35      \cr
 3626   & 0.52     & 2.63     & 0.7   &  24.6     & 19.4     &  44.3   &  0.5     &          &          &           &     &  0.25 &   0.38      \cr
 3941   & 0.67     & 2.52     & 0.9   &  21.8     & 20.1     &  35.4   &  0.6     &  16.4    & 0.9      &           &     &  0.17 &   0.35      \cr
 4245   & 0.75     & 4.24     & 1.0   &  25.9     & 27.6     &  68.5   &  0.2     &   9.1    & 0.7      &           &     &  0.20 &   0.29      \cr
 4340   & 0.54     & 4.44     & 1.0   &  41.1     & 10.9     & 131.0   &  0.1     &  36.9    & 0.5      & 11.3      & 0.7 &  0.32 &   0.48      \cr
 4596   & 0.71     & 2.78     & 1.1   &  37.5     & 33.9     &  93.0   &  0.3     &  38.4    & 0.8      &           &     &  0.13 &   0.35      \cr
 4608   & 0.72     & 3.29     & 0.9   &  36.7     & 34.1     &  62.4   &  0.3     &  24.9    & 0.9      &           &     &  0.15 &   0.40      \cr
 4643   & 0.95     & 3.06     & 0.9   &  30.6     & 38.2     &  68.8   &  0.2     &  35.8    & 0.8      &           &     &  0.17 &   0.48      \cr
\noalign{\vskip 5pt}
\noalign{\vskip 3pt\hrule\vskip 3pt}
}
\endtable

\vfill
\eject

\begintable*{8}
\caption{{\bf Table 8.} Mean B/T-ratios.}
\halign{%
\rm#\hfil&\qquad\rm#\hfil&\qquad\rm\hfil#&\qquad\rm\hfil
#&\qquad\rm\hfil#&\qquad\rm\hfil#&\qquad\rm#\hfil
&\qquad\rm\hfil#&\qquad\rm#\hfil&\qquad\hfil\rm#\cr
\noalign{\vskip 3pt\hrule\vskip 3pt}
       &             &            &        &             \cr
       & $B/T$/(N)   &$B/T$/(N)   & filter &bulge model  \cr
       & (S0)        & S0/a+Sa    &        &            \cr
\noalign{\vskip 5pt}
\noalign{\vskip 3pt\hrule\vskip 3pt}
\noalign{\vskip 5pt}

Burstein (1979)       & 0.49 $\pm$ 0.14 (11) &                    & $B$     & $R^{1/4}$   \cr
Kent (1985)           & 0.68 $\pm$ 0.18 (14) & 0.48 $\pm$ 0.25(15)& $R$     & $R^{1/4}$   \cr
Simien (1986)         & 0.57 (31)            & 0.41(5)            & $B$     & $R^{1/4}$   \cr
APB95 (1995)          & 0.32 $\pm$ 0.13 ( 7) & 0.20 $\pm$ 0.13(7) & $K$     & S\'ersic    \cr
SGA04 (2004)          & 0.63 $\pm$ 0.05 (16) &                    & $R$     & S\'ersic    \cr
CZ (2004)             & 0.45                 &                    & $R$     & $R^{1/4}$   \cr
This study            & 0.24 $\pm$ 0.11 (14) & 0.28 $\pm$ 0.14(9) & $K$     & S\'ersic    \cr
\noalign{\vskip 5pt}
\noalign{\vskip 3pt\hrule\vskip 3pt}
}
\endtable

\vfill
\eject

\psfig{file=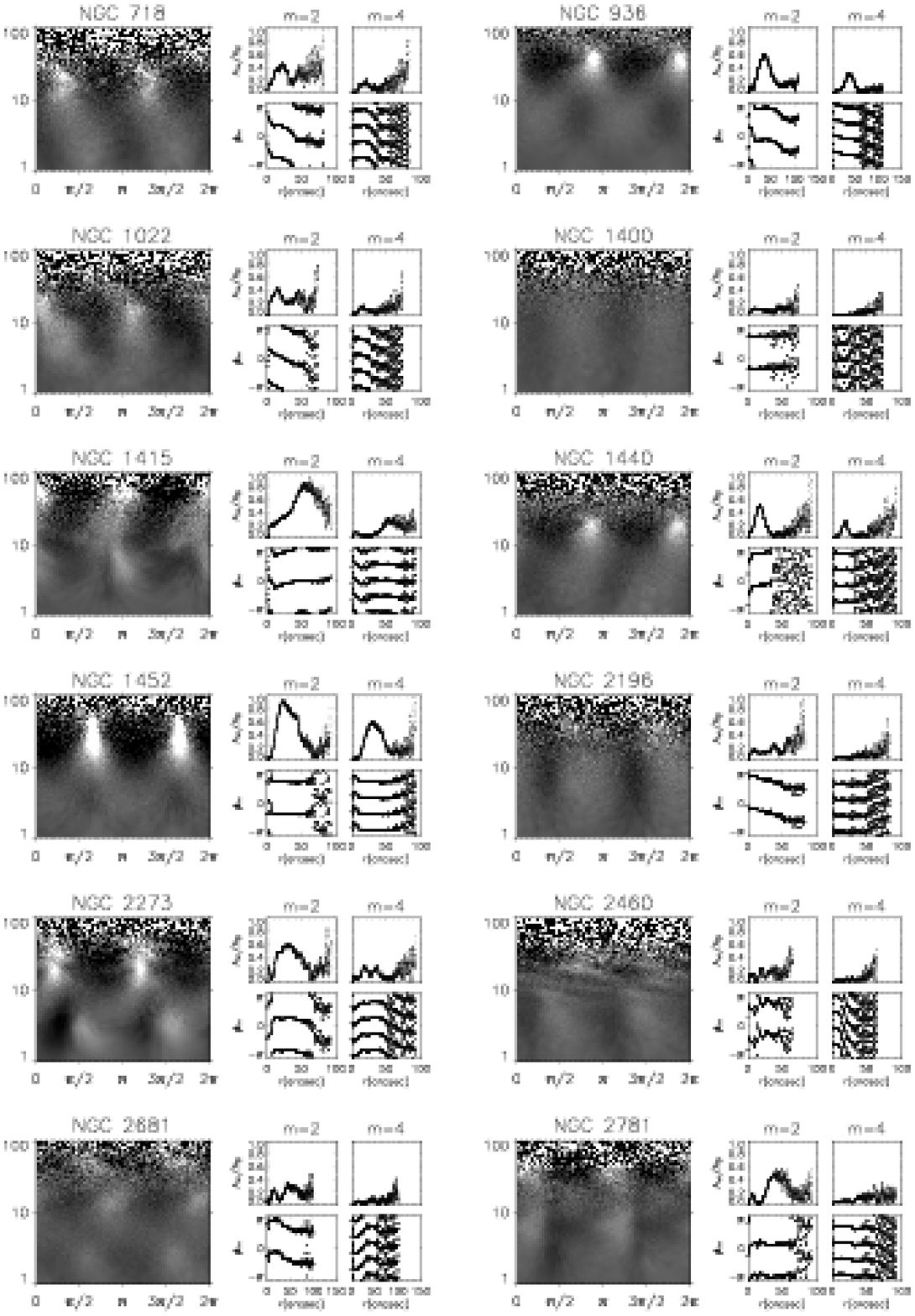,width=16cm}
Fig. 1a
\vfill
\eject

\psfig{file=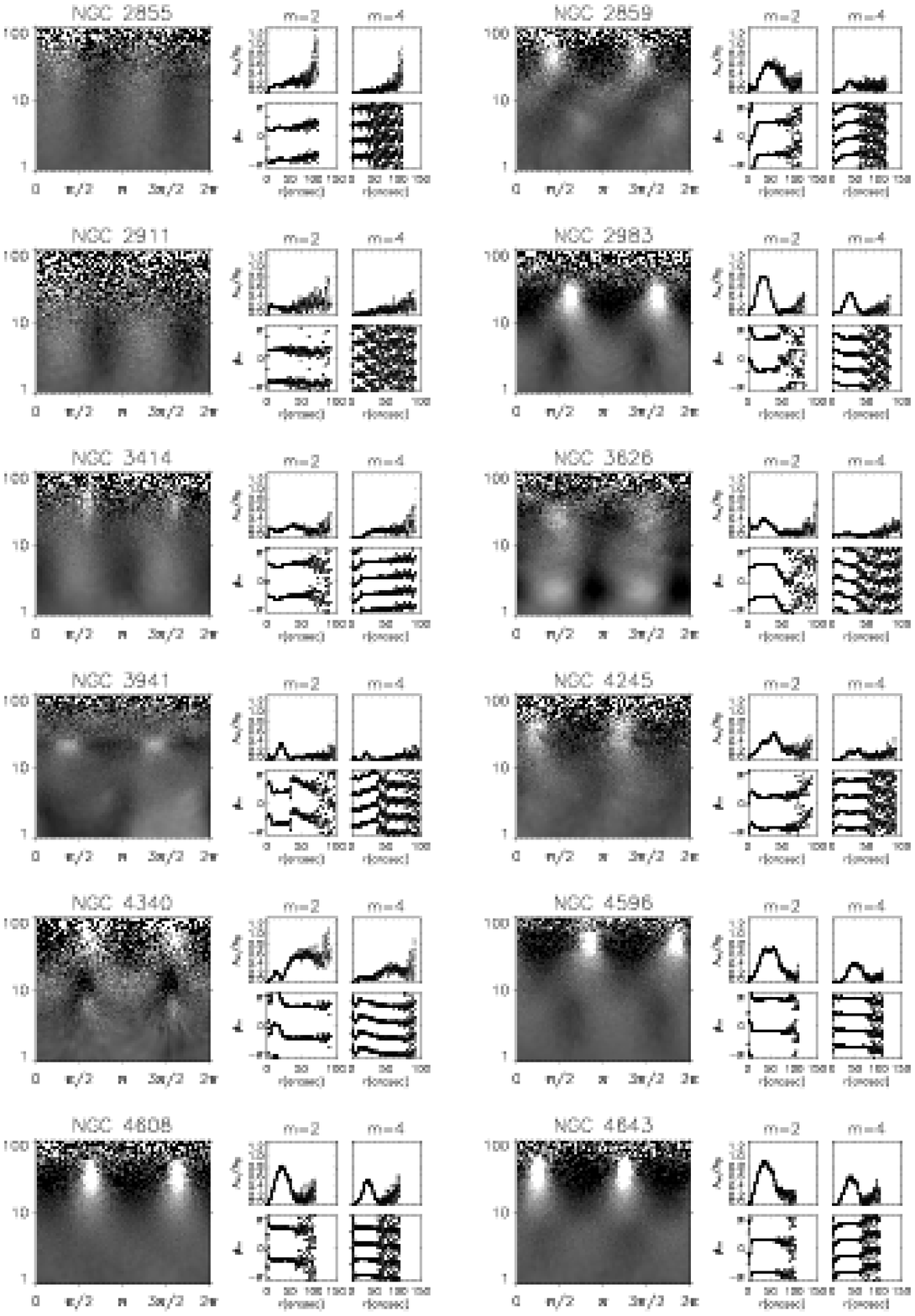,width=16cm}
Fig. 1b
\vfill
\eject

\psfig{file=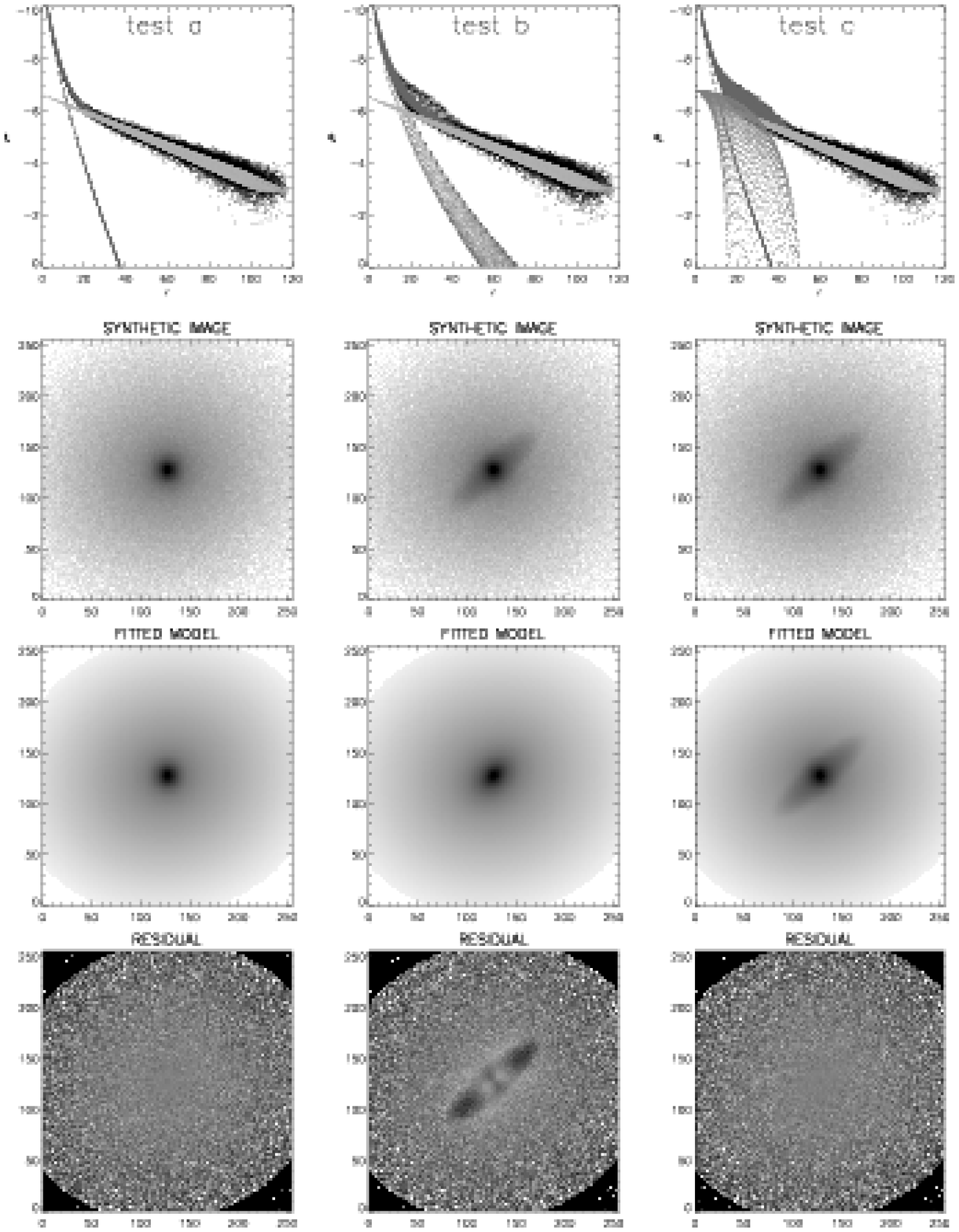,width=16cm}
Fig. 2
\vfill
\eject

\psfig{file=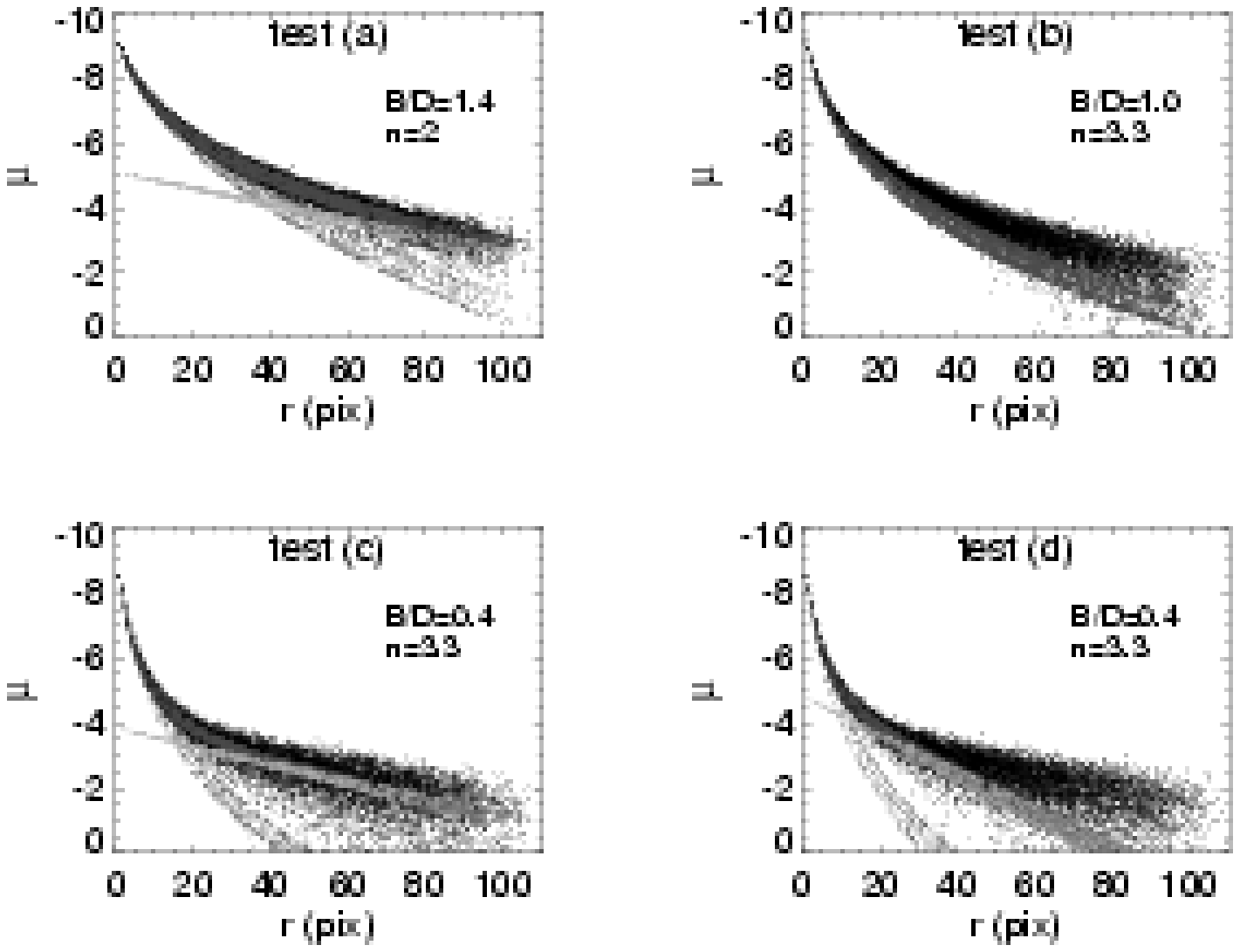,width=16cm}
Fig. 3
\vfill
\eject

\psfig{file=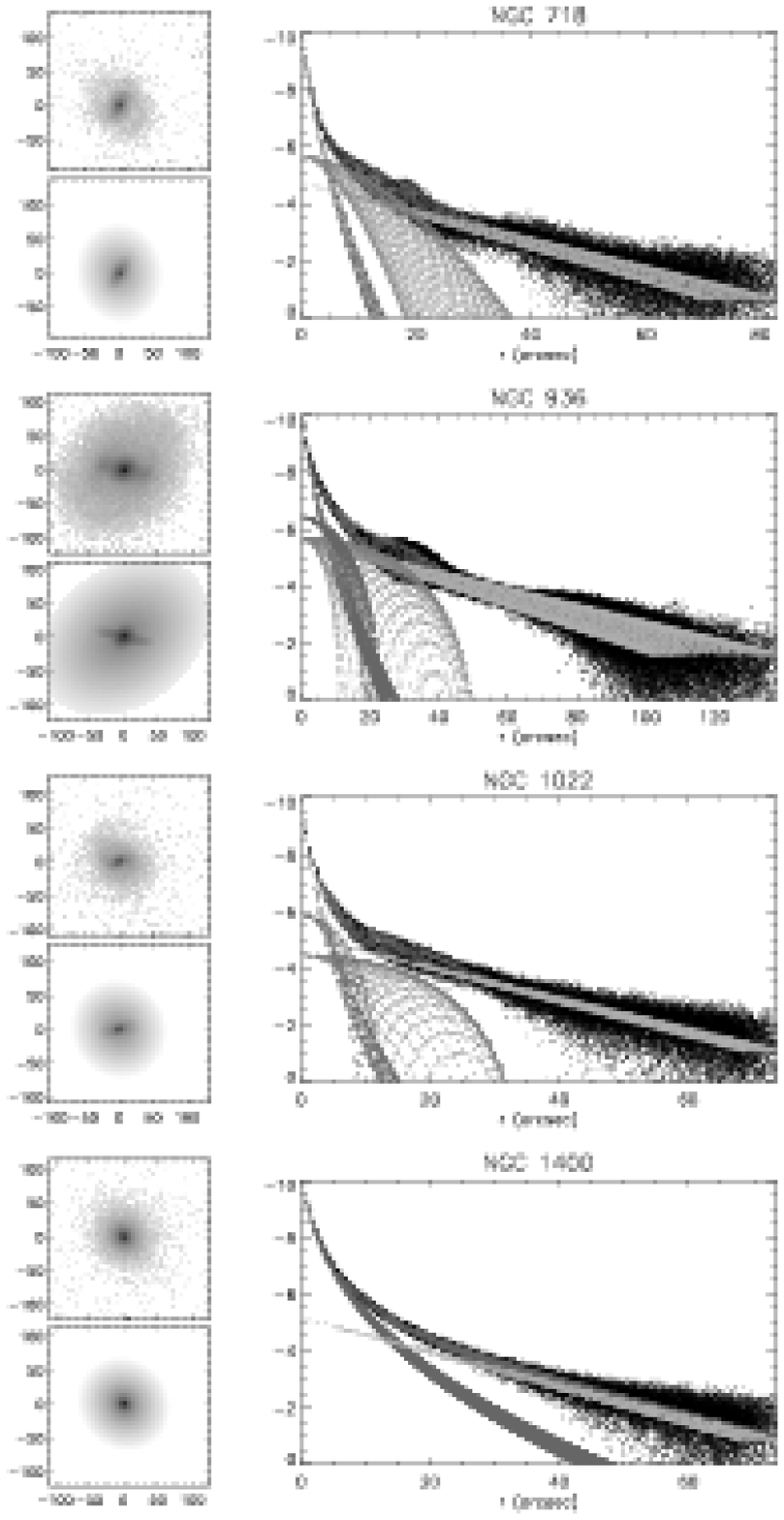,width=19cm}
Fig. 4a
\vfill
\eject

\psfig{file=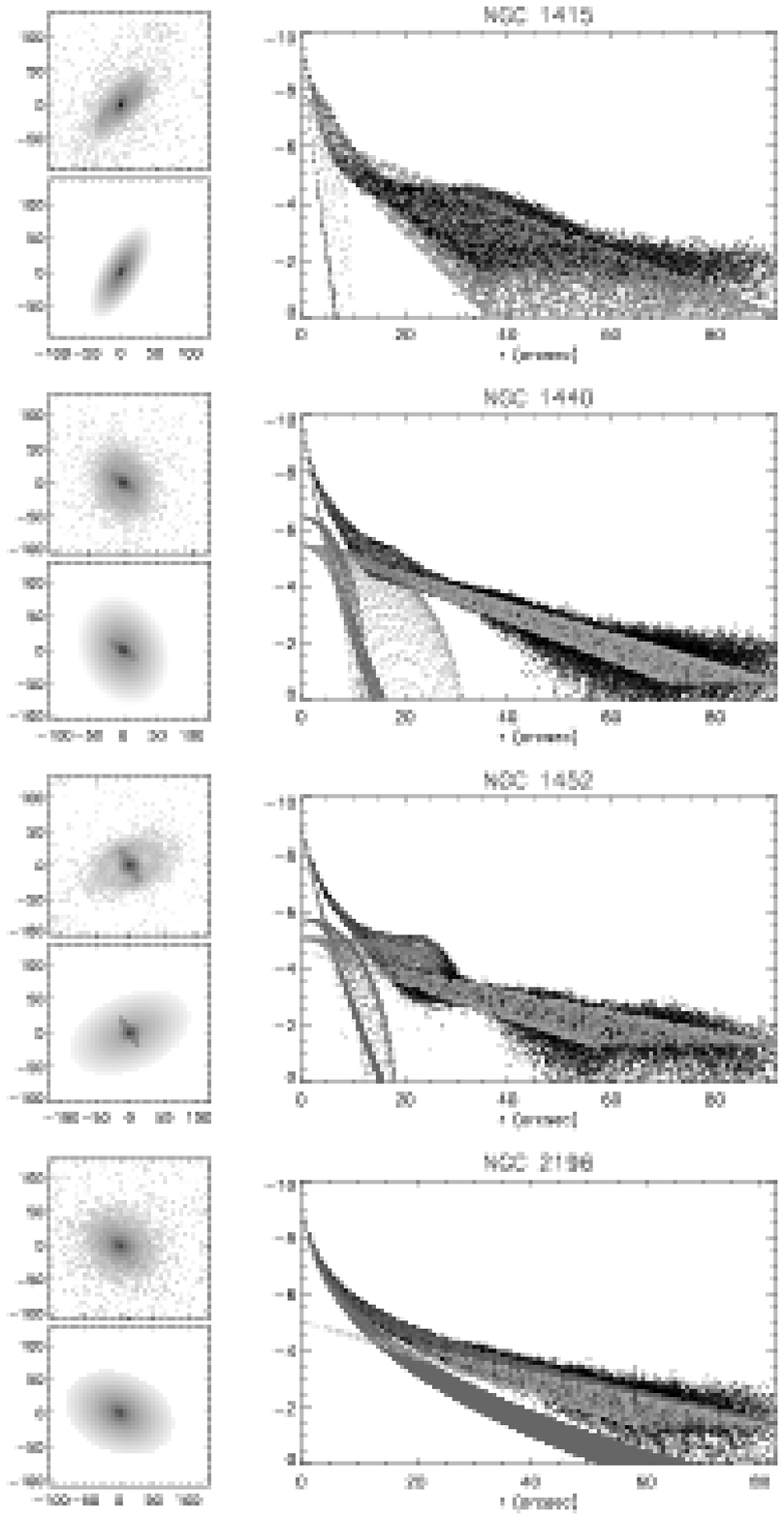,width=19cm}
Fig. 4b
\vfill
\eject

\psfig{file=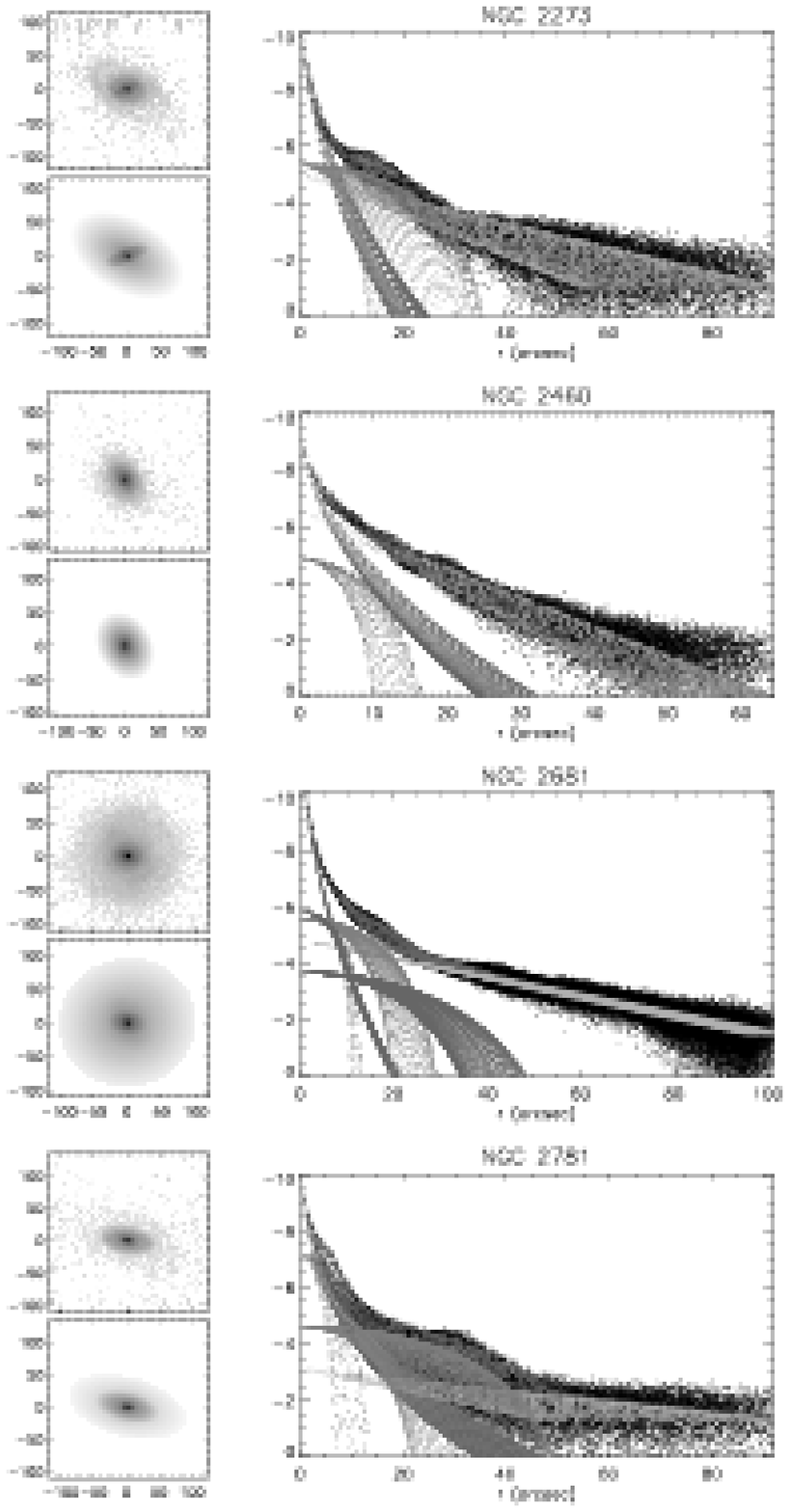,width=19cm}
Fig. 4c
\vfill
\eject

\psfig{file=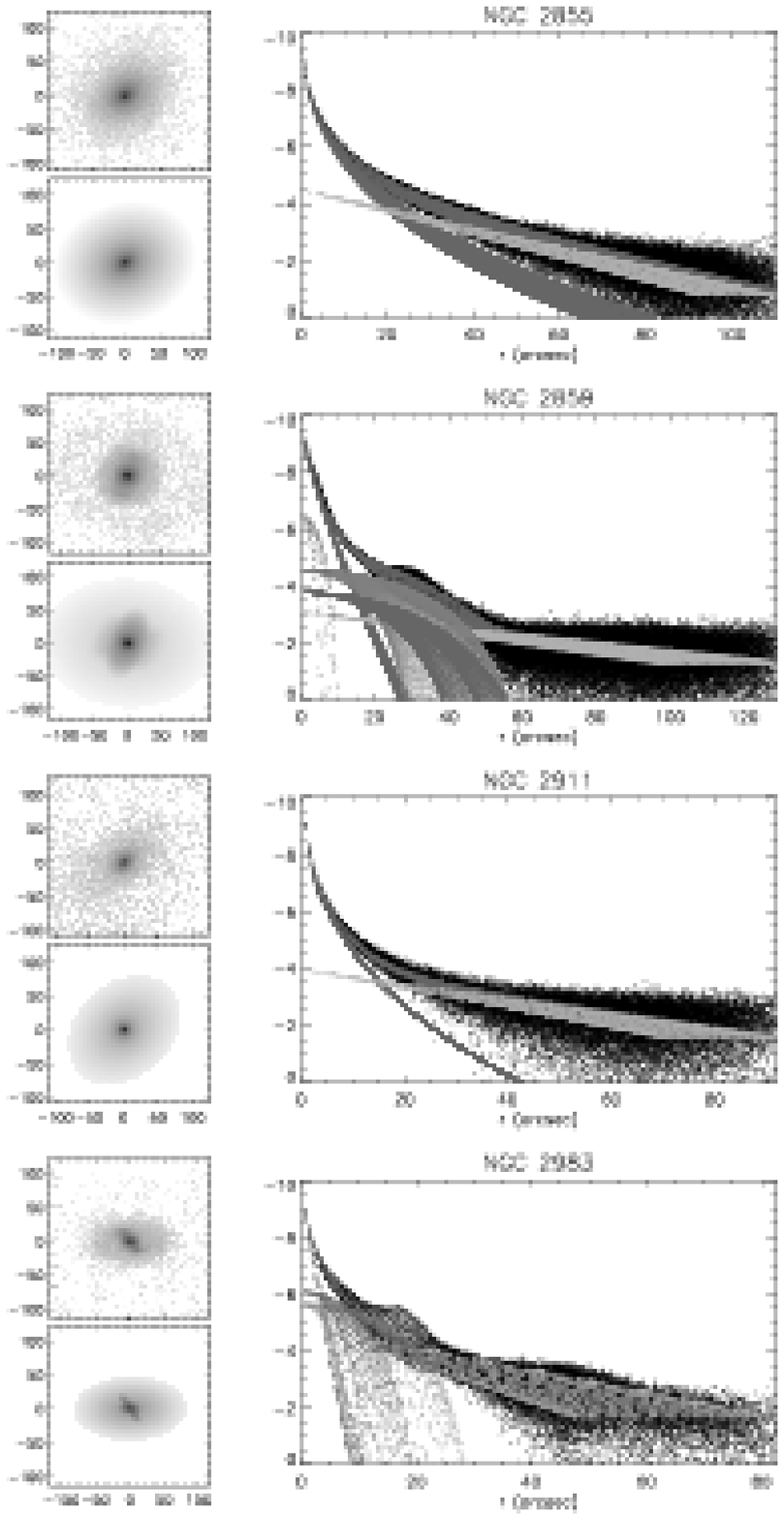,width=19cm}
Fig. 4d
\vfill
\eject

\psfig{file=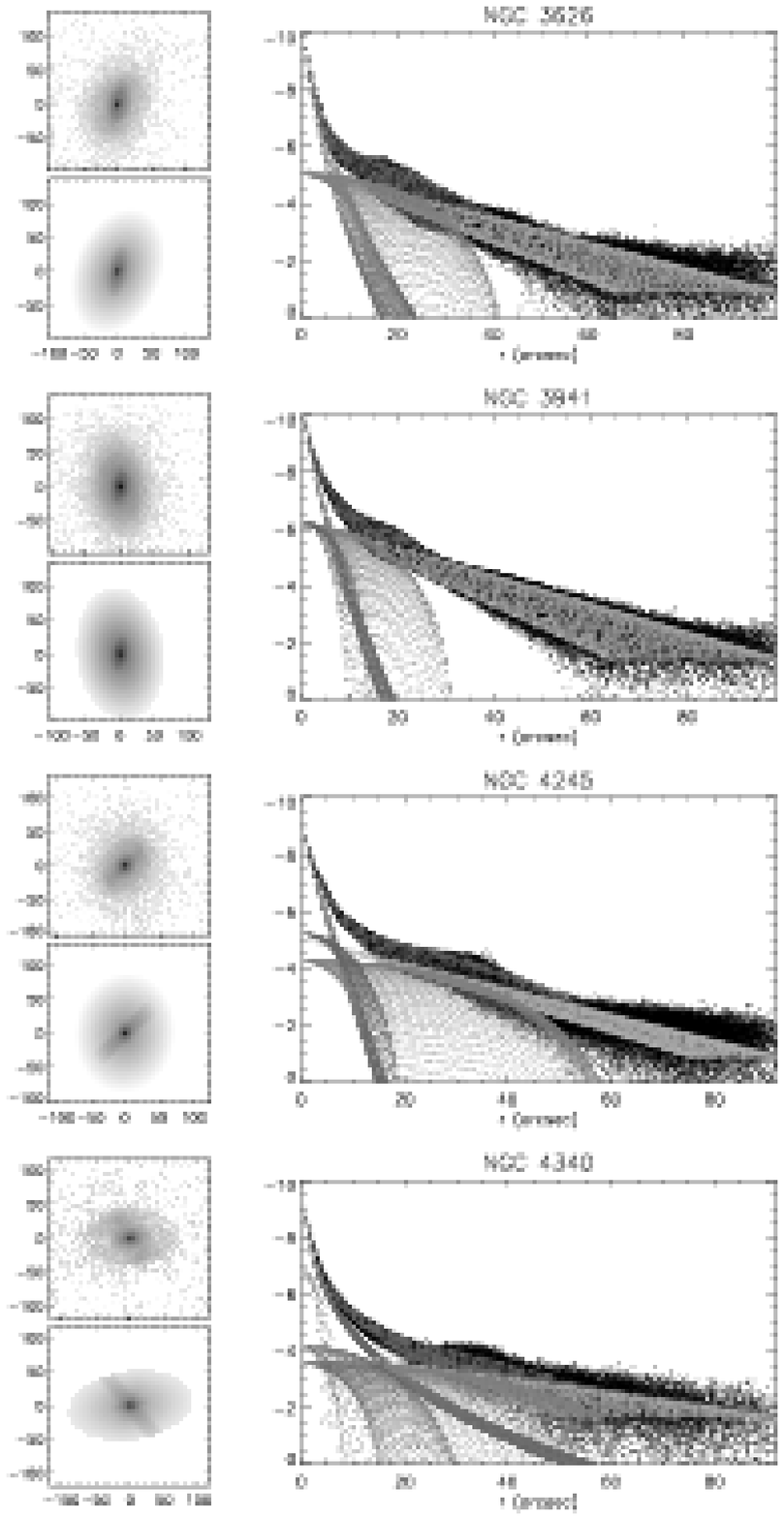,width=19cm}
Fig. 4e
\vfill
\eject

\psfig{file=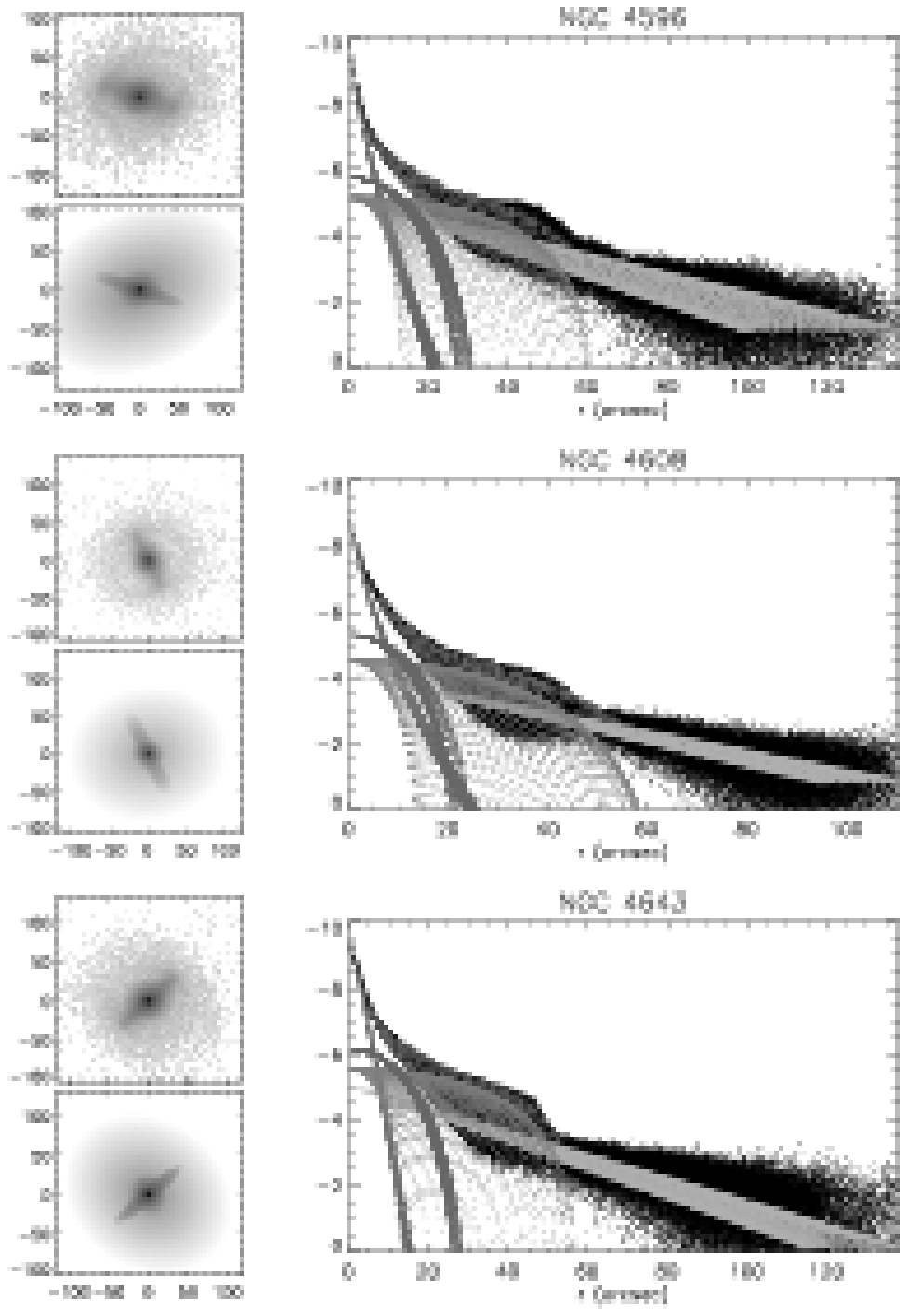,width=19cm}
Fig. 4f
\vfill
\eject

\psfig{file=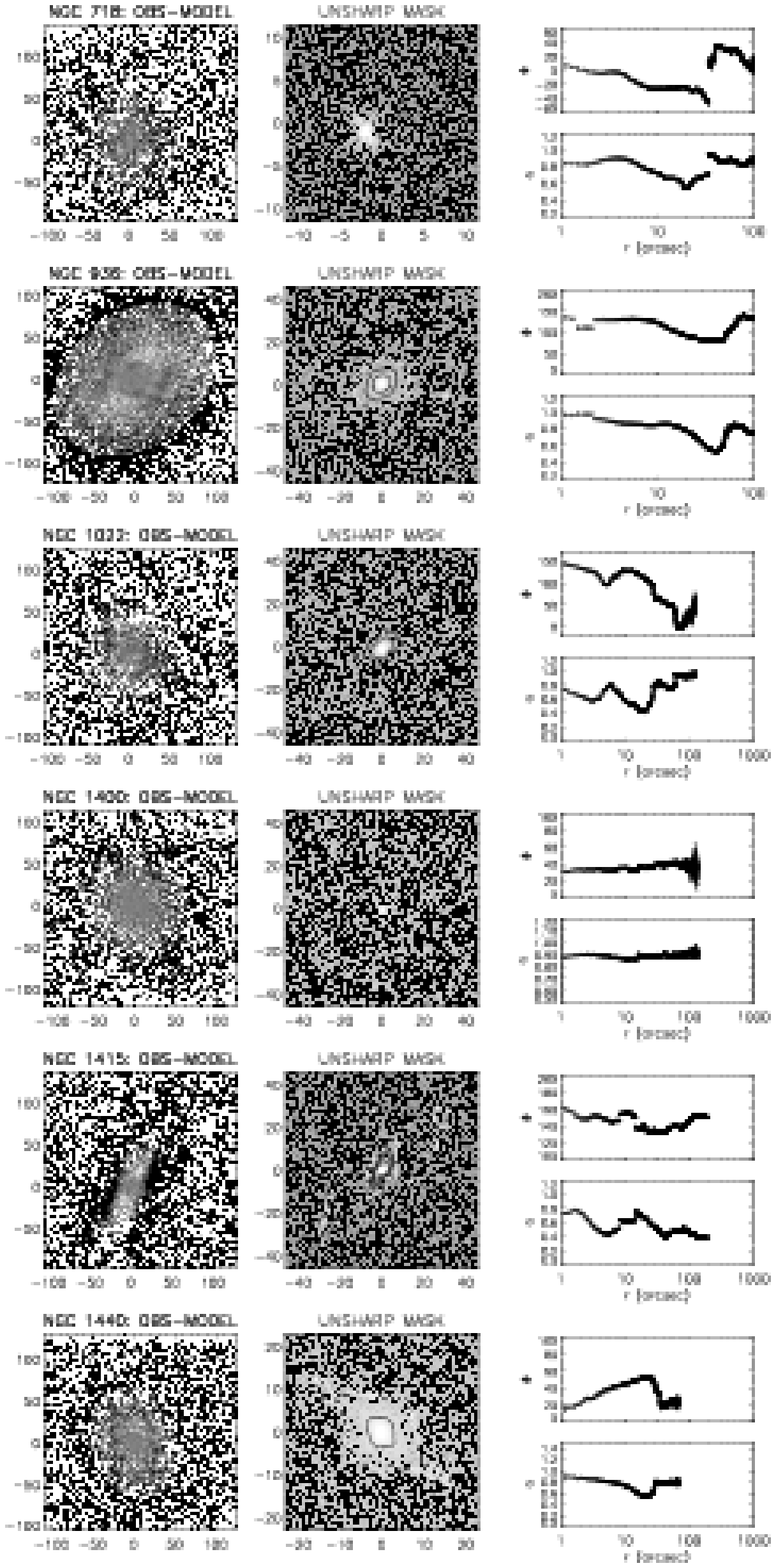,width=16cm}
Fig. 5a
\vfill
\eject

\psfig{file=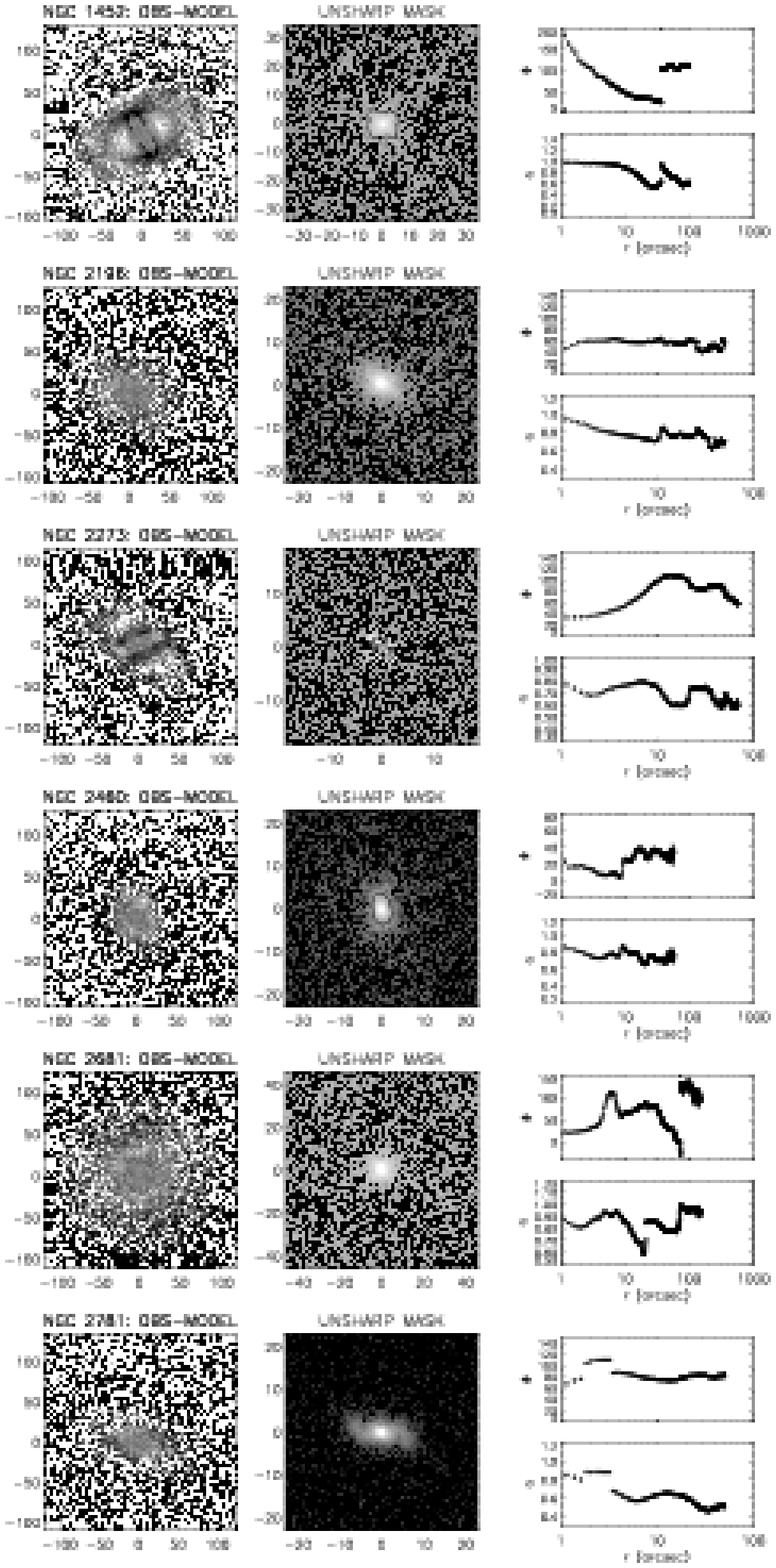,width=16cm}
Fig. 5b
\vfill
\eject

\psfig{file=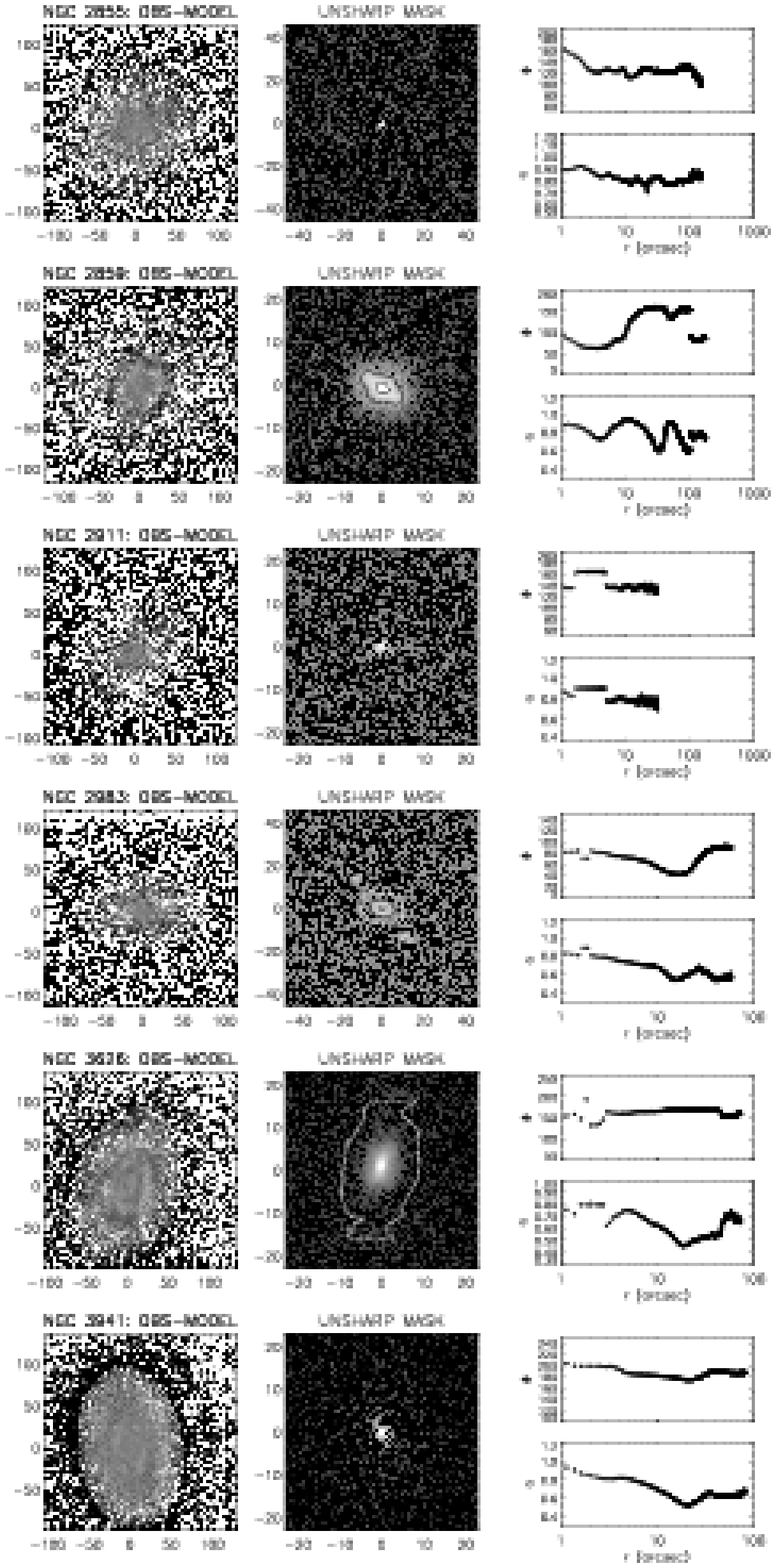,width=16cm}
Fig. 5c
\vfill
\eject

\psfig{file=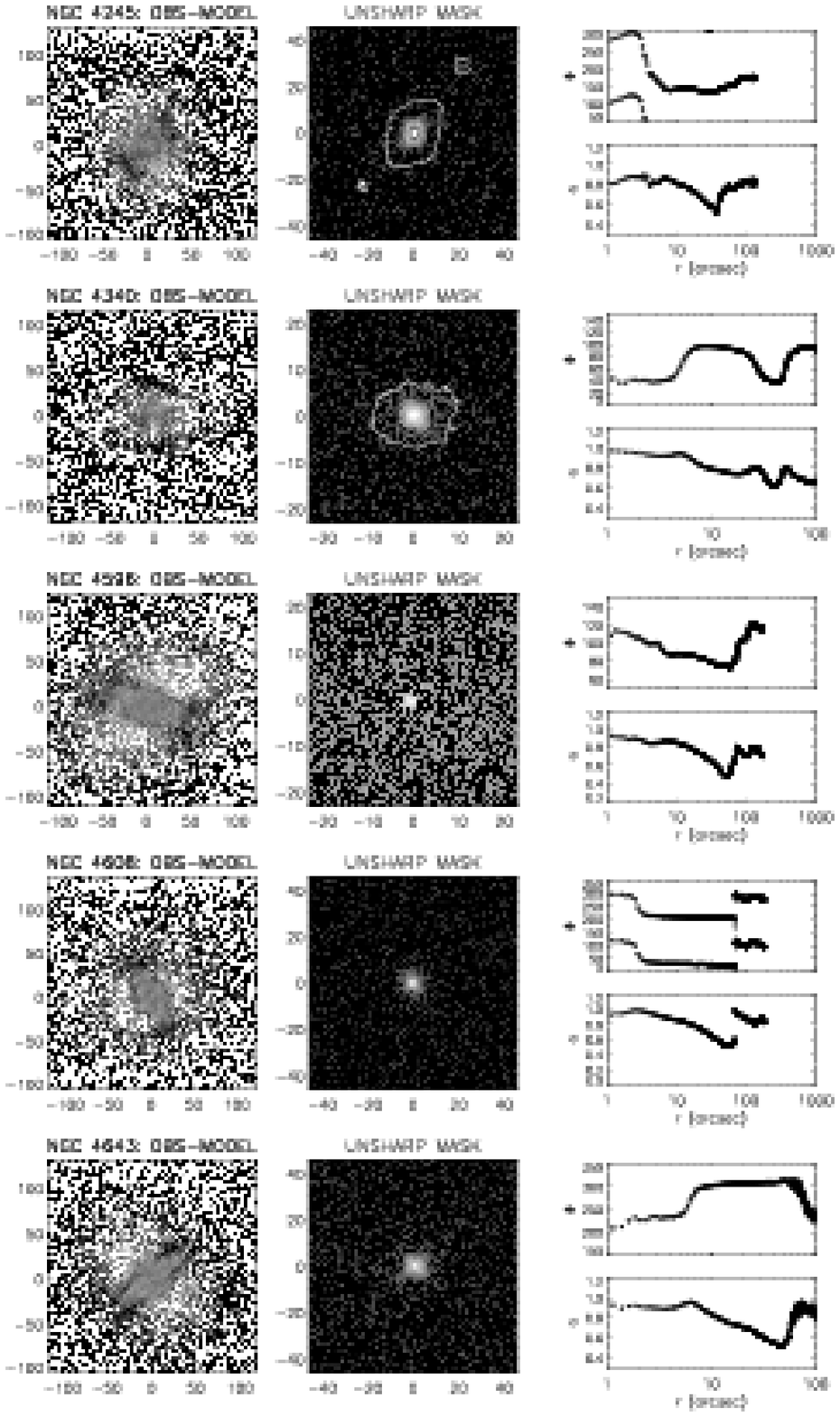,width=16cm}
Fig. 5d
\vfill
\eject

\psfig{file=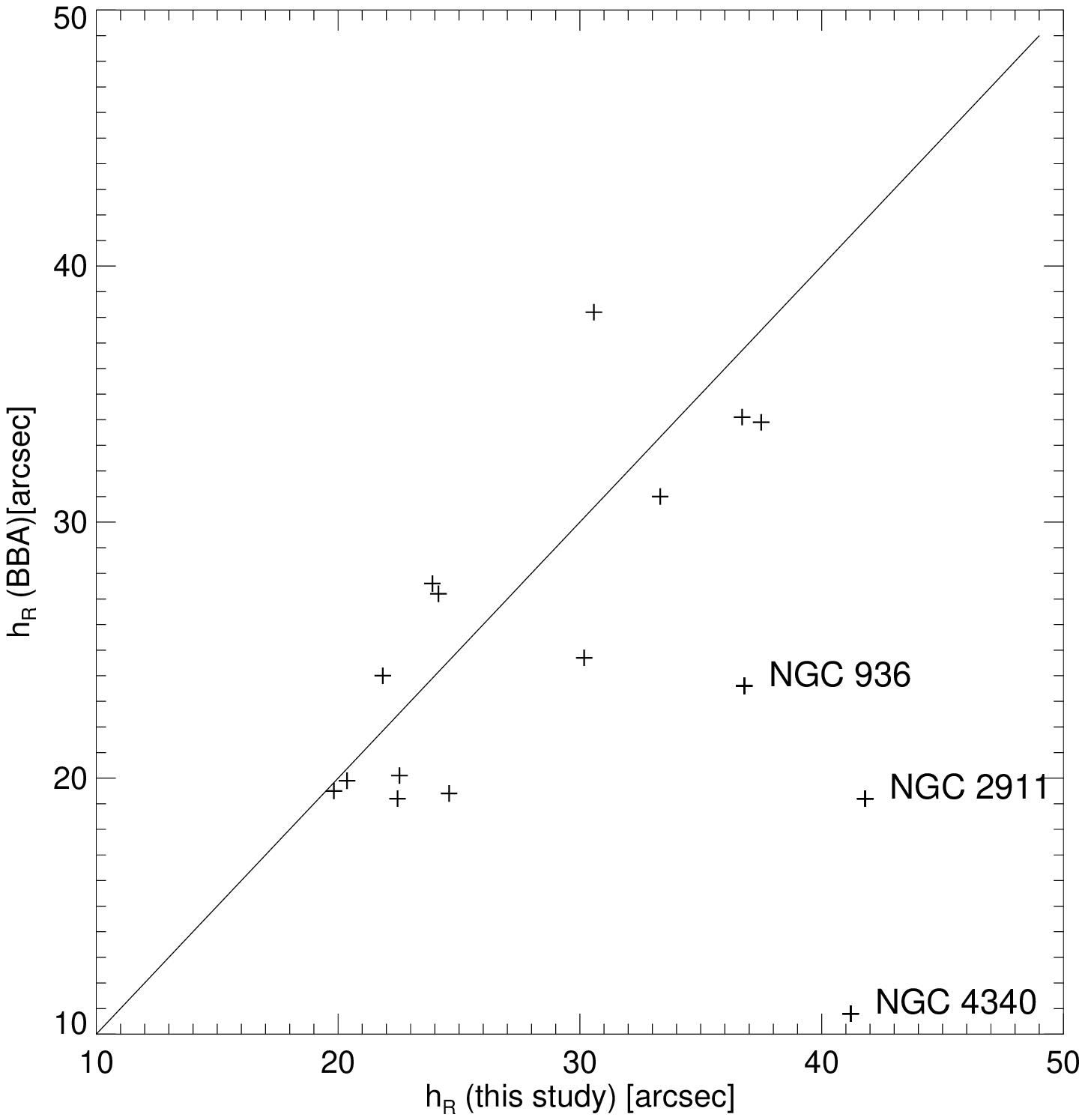,width=16cm}
Fig. 6
\vfill
\eject

\psfig{file=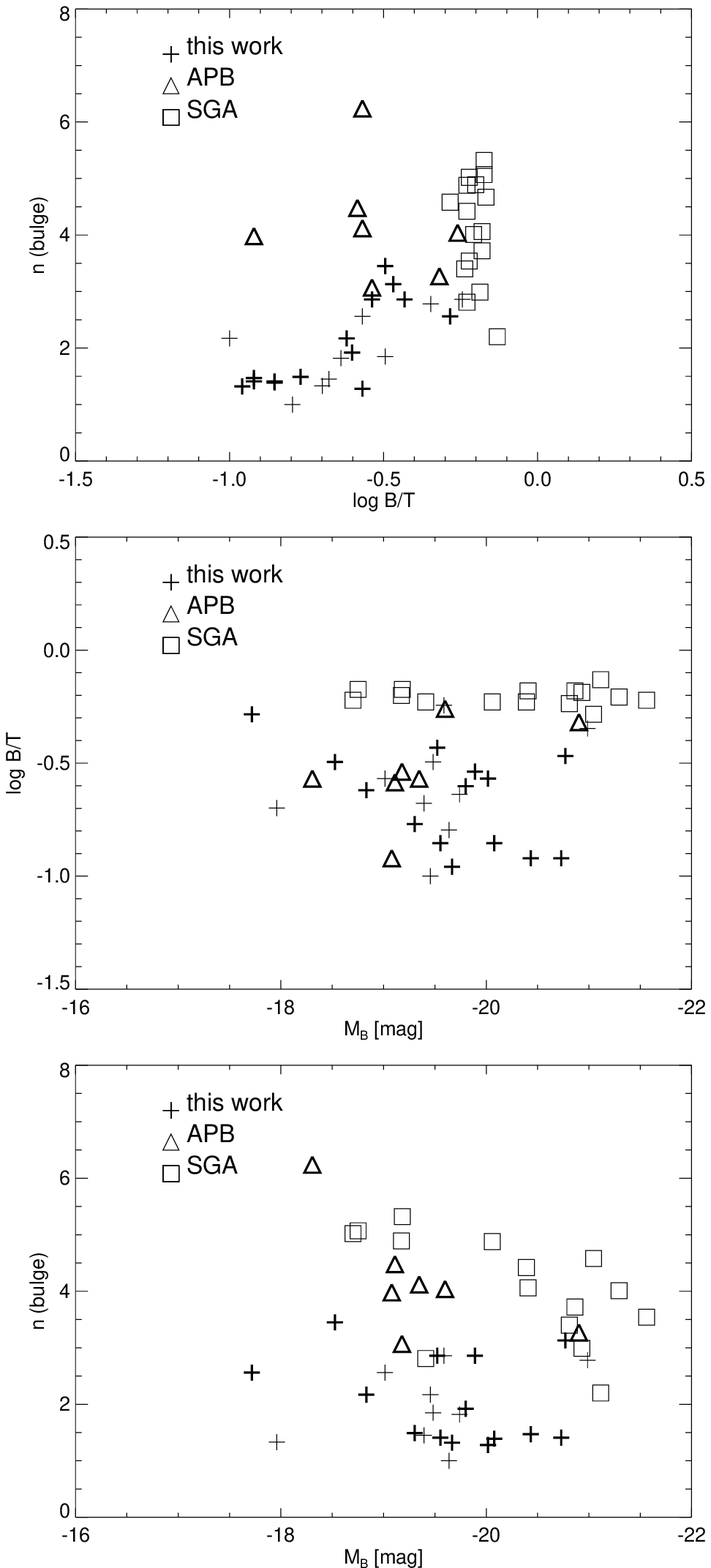,width=10cm}
Fig. 7
\vfill
\eject

\psfig{file=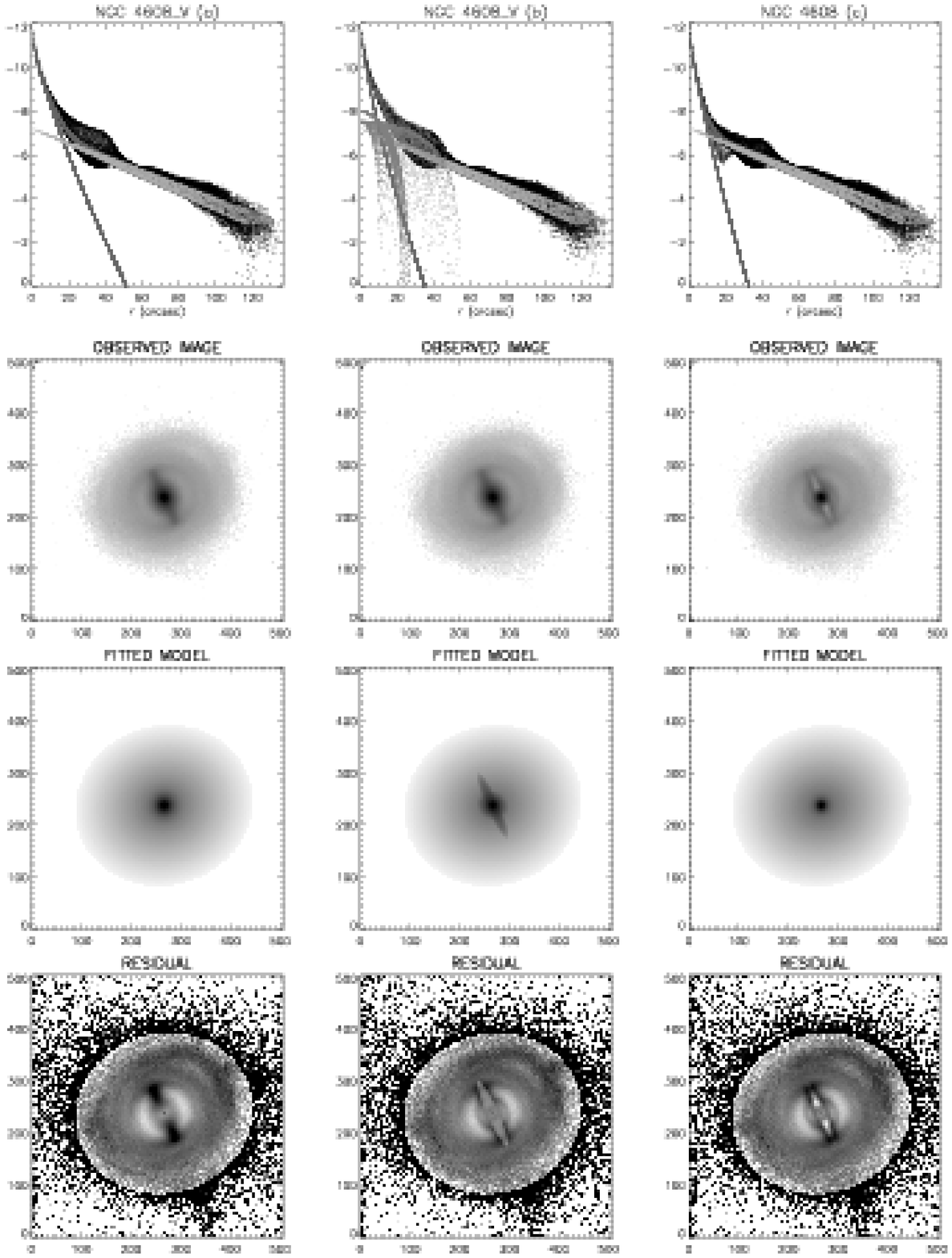,width=16cm}
Fig. 8
\vfill
\eject

\bye